\documentclass[12pt]{article}
\pdfoutput=1
\usepackage{jheppub}
\usepackage{amsmath}
\usepackage{bm}
\usepackage{slashed}
\usepackage{graphicx}
\usepackage{makecell}
\usepackage{enumitem}

\newlength{\dummysp}
\newcommand{\beq}{\begin{eqnarray}}
\newcommand{\eeq}{\end{eqnarray}}

\newcommand{\s}{{\sigma}}

\newcommand{\gappeq}{\mathrel{\rlap {\raise.5ex\hbox{$>$}}
{\lower.5ex\hbox{$\sim$}}}}
\newcommand{\lappeq}{\mathrel{\rlap{\raise.5ex\hbox{$<$}}
{\lower.5ex\hbox{$\sim$}}}}

\newcommand{\ben}{\begin{enumerate}}
\newcommand{\een}{\end{enumerate}}

\newcommand{\bit}{\begin{itemize}}
\newcommand{\eit}{\end{itemize}}

\def\[{\left [}
\def\]{\right ]}
\def\({\left (}
\def\){\right )}

\def\Z{{\mathbb Z}}

\def\a{\alpha}

\def\d{\delta}

\def\l{\lambda}

\def\s{\sigma}

\def\del{\partial}

\def\Z{\mathbb{Z}}

\def\rep{\mathcal{R}}

\def\bf{\textbf}
\def\it{\textit}
\def\f{\frac}

\def\a{\alpha}

\def\d{\delta}

\def\l{\lambda}

\def\Om{\Omega}

\def\s{\sigma}

\def\del{\partial}

\def\Z{\mathbb{Z}}

\def\rep{\mathcal{R}}

\def\bf{\textbf}
\def\it{\textit}
\def\f{\frac}

\title{Noninvertible symmetries and anomalies from gauging $1$-form electric centers}
\author{Mohamed M. Anber,}\author{Samson Y.L. Chan} 
  
\affiliation{Centre for Particle Theory, Department of Mathematical Sciences, Durham University, South Road, Durham DH1 3LE, UK}

\emailAdd{mohamed.anber@durham.ac.uk}\emailAdd{samson.y.chan@durham.ac.uk}    

\abstract{

{\flushleft{W}}e devise a general method for obtaining $0$-form noninvertible discrete chiral symmetries in $4$-dimensional $SU(N)/\mathbb Z_p$ and $SU(N)\times U(1)/\mathbb Z_p$ gauge theories with matter in arbitrary representations, where $\mathbb Z_p$ is a subgroup of the electric $1$-form center symmetry. Our approach involves placing the theory on a three-torus and utilizing the Hamiltonian formalism to construct noninvertible operators by introducing twists compatible with the gauging of $\mathbb Z_p$. These theories exhibit electric $1$-form and magnetic $1$-form global symmetries, and their generators play a crucial role in constructing the corresponding Hilbert space. The noninvertible operators are demonstrated to project onto specific Hilbert space sectors characterized by particular magnetic fluxes. Furthermore, when subjected to twists by the electric $1$-form global symmetry, these surviving sectors reveal an anomaly between the noninvertible and the $1$-form symmetries. We argue that an anomaly implies that certain sectors, characterized by the eigenvalues of the electric symmetry generators, exhibit multi-fold degeneracies. When we couple these theories to axions,  infrared axionic noninvertible operators inherit the ultraviolet structure of the theory, including the projective nature of the operators and their anomalies.  We discuss various examples of vector and chiral gauge theories that showcase the versatility of our approach.}

\begin{document}

\maketitle

\section{Introduction}

Symmetry is the bedrock upon which quantum field theory (QFT) is constructed. In the past decade, a seismic shift has occurred in our understanding of symmetries, transcending their conventional application to mere point-like particles. In the contemporary paradigm, a $p$-form symmetry in $4$ dimensions is linked to operators residing on $(3-p)$-dimensional topological manifolds that act on $p$-dimensional objects charged under the symmetry. Moreover,  over the last couple of years, symmetries have expanded their domain to encompass operators that defy the conventional notion of inversion. These are known as noninvertible symmetries. While noninvertible symmetries initially found their roots and applications in the realm of $2$-dimensional QFT, see, e.g., \cite{Fuchs:2002cm, Komargodski:2020mxz}, their significance in the context of 4-dimensional QFT sparked a deluge of research endeavors in this area (a non-comprehensive list is \cite{Nguyen:2021yld,Nguyen:2021naa,Choi:2021kmx,Wang:2021vki,Bhardwaj:2022yxj,
Choi:2022zal,Kaidi:2022uux,Choi:2022jqy,Cordova:2022ieu,Choi:2022rfe,Bartsch:2022mpm,Heckman:2022muc,Cordova:2022fhg,Karasik:2022kkq,GarciaEtxebarria:2022jky,Choi:2022fgx,Yokokura:2022alv,Bhardwaj:2022kot,Bhardwaj:2022maz,Bartsch:2022ytj,Heckman:2022xgu,Apte:2022xtu,Delcamp:2023kew,Kaidi:2023maf,Putrov:2023jqi,Dierigl:2023jdp,Choi:2023pdp}. Also, see \cite{Schafer-Nameki:2023jdn,Shao:2023gho} for reviews.)

It is well known that quantum electrodynamics has a classical $U(1)_\chi$ axial symmetry that breaks down because of the Adler–Bell–Jackiw (ABJ)  anomaly. However,  it was realized in \cite{Choi:2022jqy,Cordova:2022ieu} that the axial symmetry does not completely disappear. Instead, it resurfaces as a noninvertible symmetry for each fractional element of the classical $U(1)_\chi$.  This profound reinterpretation of symmetries prompted a compelling quest to unearth analogous structures in QFT. In \cite{Anber:2023pny}, one of the authors established a technique for unveiling noninvertible $0$-form symmetries within $SU(N)\times U(1)$ gauge theories in the presence of matter in representation ${\cal R}$. This approach employed the Hamiltonian formalism, where the theory was put on a three-dimensional torus $\mathbb T^3$, subjecting it to $\mathbb Z_N$ magnetic twists along all three spatial directions. Taking the matter to be a single Dirac fermion, this theory is endowed with invertible $\mathbb Z_{\scriptsize 2\mbox{gcd}(T_{\cal R}, d_{\cal R})}^\chi$ $0$-form chiral symmetry, where $T_{\cal R}$ and $d_{\cal R}$ are the Dynkin index and dimension of ${\cal R}$, respectively. Yet, it was shown that the theory also possesses a noninvertible $\tilde {\mathbb Z}_{2T_{\cal R}}^\chi$ $0$-form chiral symmetry\footnote{In this paper, a tilde is used to indicate that a given symmetry or operator is noninvertible.}. Such symmetry acts on the Hilbert space projectively by selecting special sectors characterized by certain magnetic numbers. 
New noninvertible symmetries were also revealed in \cite{Argurio:2023lwl} in theories with mixed anomalies between $\mathbb Z_2^{(1)}$ $1$-form and $0$-form discrete chiral symmetries. 

The topological essence of symmetries, encompassing the noninvertible variants, underscores their sensitivity to the global structure of the gauge group. Consequently, the inquiry arises: how do we identify these noninvertible symmetries within a general gauge group, characterized as either $SU(N)/\mathbb Z_p$ or $SU(N)\times U(1)/\mathbb Z_p$ where $\mathbb Z_p$ is a subgroup of the center symmetry? In this work, we answer this question by devising a general method that applies to any theory with a direct multiplication of abelian and semi-simple nonabelian gauge groups quotiented by a discrete center, whether the theory is vector-like or chiral. This is achieved by putting the theory on $\mathbb T^3$ and turning on magnetic fluxes in a refined subgroup of $\mathbb Z_N$, depending on the matter content as well as the global structure of the gauge group.  

In the context of $SU(N)$ gauge theory, the introduction of matter characterized by an $N$-ality $n$ has the effect of breaking the $\mathbb Z_N$ center of the group down to a subgroup $\mathbb Z_q$, where $q$ is the greatest common divisor (gcd) of $N$ and $n$. Our focus is on understanding the noninvertible $0$-form symmetries present in the $SU(N)/ \mathbb Z_p$ gauge theories, where $\mathbb Z_p$ is a subgroup of the remaining center $\mathbb Z_q$. These theories exhibit both electric $\mathbb Z_{q/p}^{(1)}$ and magnetic $\mathbb Z_p^{(1)}$ 1-form global symmetries\footnote{There are $p$ distinct theories $\left(SU(N)/\mathbb Z_p\right)_n$, where $n=0,1,..,p-1$ are the discrete $\theta$-like parameters \cite{Aharony:2013hda}. These theories differ by the set of compatible line operators (Wilson, 't Hooft, and dyonic operator). Here, we restrict our analysis to $n=0$.}.
To identify the noninvertible symmetries, we initiate the process starting from $SU(N)$ theory endowed with a single Dirac fermion in representation ${\cal R}$, which possesses an invertible $\mathbb Z_{2T_{\cal R}}^\chi$ chiral symmetry.  We then subject this theory to electric and magnetic twists characterized by elements of $\mathbb Z_p$. If the theory exhibits a mixed anomaly between its chiral and electric $\mathbb Z_p^{(1)}$ 1-form symmetries, the act of gauging $\mathbb{Z}_p$ effectively reveals the chiral symmetry as noninvertible. The construction of a gauge-invariant operator corresponding to the noninvertible symmetry $\tilde {\mathbb Z}_{2T_{\cal R}}^\chi$ involves several steps. First, we create a topological operator by integrating the anomalous current conservation law over  $\mathbb T^3$. The resulting operator is not invariant under  $\mathbb Z_p$ gauge transformations. Yet, we can restore gauge invariance by summing over all possible $\mathbb Z_p$ gauge-transformed operators. This process results in a noninvertible chiral symmetry operator that projects onto specific sectors in the Hilbert space, each characterized by certain 't Hooft lines charged under the magnetic $\mathbb Z_p^{(1)}$ 1-form symmetry. $\tilde {\mathbb Z}_{2T_{\cal R}}^\chi$ can exhibit further anomalies when subjected to twists by the electric $\mathbb Z_{q/p}^{(1)}$ $1$-form symmetry, implying that states within the Hilbert space of the $SU(N)/\mathbb Z_p$ gauge theory will display multiple degeneracies.

We employ a similar approach to identify noninvertible symmetries in $SU(N)\times U(1)/\mathbb Z_p$ gauge theories, where $\mathbb Z_p$ is a subgroup of the electric $\mathbb Z_N^{(1)}$ 1-form center symmetry. Unlike in $SU(N)$ theories, the introduction of matter does not reduce the $\mathbb Z_N$ center. This is due to the presence of an abelian $U(1)$ sector, which ensures that all matter representations adhere to the cocycle condition. In addition to the $1$-form electric center symmetry, this theory is also endowed with a magnetic $U^{(1)}_m(1)$ 1-form symmetry.  $SU(N)$ gauge theory with matter exhibits an anomaly between its chiral and $U(1)$ baryon-number symmetries. Gauging the latter transforms the theory into an $SU(N)\times U(1)$ gauge theory and reveals the chiral symmetry $\tilde {\mathbb Z}_{2T_{\cal R}}^\chi$  as noninvertible. Placing the theory on $\mathbb T^3$ enables us to construct the corresponding noninvertible chiral operator by summing over large $U(1)$ gauge transformations with distinct winding numbers. Furthermore, since the theory exhibits a $1$-form electric center symmetry, we can decorate the noninvertible operator with $\mathbb Z_N$ magnetic twists. If we choose to further gauge a $\mathbb Z_p^{(1)}$ subgroup of the electric $\mathbb Z_N^{(1)}$ symmetry, thereby resulting in the $SU(N)\times U(1)/\mathbb Z_p$ theory, we must ensure that the noninvertible operator remains invariant under $\mathbb Z_p$ gauge transformation. This is accomplished by summing over all $\mathbb Z_p$ gauge-transformed chiral operators. Once again, we discover that the resultant operator projects onto specific sectors within the Hilbert space, distinguished by the presence of 't Hooft lines charged under $U^{(1)}_m(1)$. The noninvertible symmetry also exhibits a mixed anomaly with the remaining electric $\mathbb Z_{N/p}^{(1)}$ global symmetry. The anomaly implies that certain sectors of the theory, designated by certain $\mathbb Z_{N/p}^{(1)}$ electric fluxes, exhibit multi-fold degeneracy.  

Placing the theory on $\mathbb T^3$ offers a distinct advantage: it presents a systematic approach for computing the 't Hooft anomalies inherent to a given theory. Simultaneously, it provides a means to construct the Hilbert space explicitly. In our work, we put a significant emphasis on this Hilbert space construction, shedding light on the intricate relationship between Wilson's lines, 't Hooft lines, and the noninvertible operator. Specifically, through several illustrative examples, we showcase how the noninvertible chiral operator, within the framework of the Hilbert space and Hamiltonian formalism, acts to annihilate the minimal 't Hooft lines.
 
We also introduce couplings of gauge theories to axions. The underlying renormalization group invariance of the noninvertible symmetries, along with their associated anomalies, guarantees that the infrared (IR) axion physics faithfully inherits all the characteristics of the theory at the ultraviolet (UV) level. We substantiate this by explicitly constructing noninvertible chiral operators, commencing from the IR anomalous axion current conservation law. In our exploration, we offer concrete illustrations of various UV theories and their corresponding IR axion physics manifestations.

 This paper is organized as follows. In Section \ref{Preleminaries}, we provide a concise overview of the essential elements required for the development of noninvertible symmetries. This section encompasses the introduction of our notation, a review of the path-integral formalism on the 4-torus ($\mathbb{T}^4$), 't Hooft twists, and the Hamiltonian formalism on $\mathbb{T}^3$.
Moving on to Section \ref{SUNMATHBB ZP THEORIES}, we proceed to construct noninvertible symmetries within the context of $SU(N)/\mathbb{Z}_p$ theories while also identifying their associated anomalies. This section concludes with the presentation of specific examples of noninvertible symmetries in both vector and chiral gauge theories.
In Section \ref{SUNTIMES U1MODZP}, we replicate the same analysis, this time focusing on $SU(N)\times U(1)/\mathbb{Z}_p$. Two examples are discussed, including the Standard Model (SM), and we demonstrate that the SM lacks noninvertible symmetries within its non-gravitational sector.
Finally, our paper culminates in Section \ref{Coupling gauge theories to axions and noninvertible symmetries}, where we explore the coupling of gauge theories to axions. We show that noninvertible symmetry operators can also be constructed using the axion anomalous current.

\section{Preliminaries}
\label{Preleminaries}
In this section, we review the path integral and the Hamiltonian formalisms of gauge theories put on a compact manifold with possible 't Hooft twists, both in space and time directions. Additionally, we examine the global symmetries and anomalies in both formalisms, providing an exploration of these key aspects. We base our formalism and notation on \cite{tHooft:1981sps,vanBaal:1984ra,Cox:2021vsa,Anber:2023pny},  and set the stage for constructing the noninvertible operators we carry out in the subsequent sections.  While some results in this section are new,  many are a mere review of previous results. Moreover, some details are avoided, referring the reader to the literature for an in-depth discussion. Yet, the information encapsulated here is necessary to make this paper self-contained. 

\subsection{Twisting in the Path integral}

\subsubsection*{Pure $SU(N)$ theory}

We begin by reviewing 't Hooft twists on a compact $4$-dimensional Euclidean manifold with nontrivial $2$-cycles.  We consider $SU(N)$ pure Yang-Mills (YM) theory on $\mathbb T^4$, where $\mathbb T^4$ is a $4$-torus with periods of length $L_\mu$, $\mu=1,2,3,4$\footnote{YM theory on $\mathbb T^4$ with $\mathbb Z_N$ 't Hooft twists dates back to the original work by 't Hooft \cite{tHooft:1981sps}. The groundbreaking paper \cite{Gaiotto:2017yup} unveiled a novel mixed anomaly, specifically involving the electric $\mathbb Z_N^{(1)}$ $1$-form symmetry. The background of the $1$-form symmetry is a $2$-form field that can be implemented via a 't Hooft twist. This fact led to a wave of enthusiasm to understand the semi-classical limit of gauge theories on $\mathbb T^4$ or $\mathbb T^2\times \mathbb R^2$ \cite{Anber:2022qsz,Anber:2023sjn,Poppitz:2022rxv,Tanizaki:2022ngt}.}. The $SU(N)$ gauge fields $A_\mu$ are taken to obey the boundary conditions
\begin{eqnarray}
A_\nu(x+L_\mu \hat e_\mu)=\Omega_\mu\circ A_\nu(x)\equiv\Omega_\mu(x) A_\nu(x) \Omega_\mu^{-1}(x)-i \Omega_\mu(x) \partial_\nu \Omega_\mu^{-1}(x)\,,
\label{conditions on gauge field}
\end{eqnarray}
upon traversing $\mathbb T^4$ in each direction. The transition functions $\Omega_\mu$ are $N\times N$ unitary matrices in the defining representation of $SU(N)$, and  $\hat e_\nu$ are unit vectors in the $x_\nu$ direction. The subscript $\mu$ in $\Omega_\mu$ means that the function $\Omega_\mu$ does not depend on the coordinate $x_\mu$. Then, the compatibility of (\ref{conditions on gauge field}) at the corners of the $x_\mu-x_\nu$ plane of $\mathbb T^4$ gives the cocycle condition
\begin{eqnarray}\label{cocycle}
\Omega_\mu (x + \hat{e}_\nu L_\nu) \; \Omega_\nu (x) = e^{i {2 \pi n_{\mu\nu} \over N}} \Omega_\nu (x+ \hat{e}_\mu L_\mu) \; \Omega_\mu (x)\,.
\end{eqnarray}
The exponent $e^{i {2 \pi n_{\mu\nu} \over N}}$, with anti-symmetric integers $n_{\mu\nu}=-n_{\nu\mu}$, is the $\mathbb Z_N$ center of $SU(N)$. The freedom to twist by elements of the center stems from the fact that both the transition function and its inverse appear in (\ref{conditions on gauge field}).  This is also equivalent to the fact that the Wilson lines in pure $SU(N)$ gauge theory are charged under the electric $\mathbb Z_N^{(1)}$ $1$-form center symmetry. The fundamental (defining representation) Wilson lines wind around the $4$ cycles and are given by
\begin{eqnarray}
W_\mu=\mbox{tr}_\Box\left[P e^{i \int_{x_\mu=0}^{x_\mu=L_\mu} A_\mu}\Omega_{\mu}\right]\,,
\label{Wilson lines}
\end{eqnarray}
where $\Box$ denotes the defining representation of $SU(N)$
and the insertion of the transition function $\Omega_\mu$ ensures the gauge invariance of the lines. It will be useful to break $n_{\mu\nu}$ into spatial (magnetic) $m_i$ and temporal (electric) $k_i$ twists:
\begin{eqnarray}\label{electric and magnetic fluxes def}
k_i\equiv n_{i4}\,,\quad n_{ij}\equiv \epsilon_{ijk}m_k\,,
\end{eqnarray}
and $i,j=1,2,3$ or $x,y,z$. We also use bold-face letters, e.g., $\bm k\equiv(k_1,k_2,k_3)$, to denote $3$-dimensional vectors.  When applied to the gauge fields on $\mathbb T^4$, the twists induce a background with fractional topological charge\footnote{\label{t hooft flux on T4}The simplest way to find the topological charge is by activating the electric and magnetic 't Hooft fluxes along the Cartan generators of $SU(N)$; see, e.g., \cite{Anber:2019nze}. We set $F_{12}=-\frac{2\pi m_3} {L_1L_2} \nu_aH_a$ and $F_{34}=\frac{2\pi k_3 }{L_3L_4}\nu_aH_a$ along the $1$-$2$ and $3$-$4$ planes (and similar expressions in the rest of the planes), where $H_a$ are the Cartan generators, $\nu_a$ are the weights of the defining representation,  $a=1,..., N-1$, with summation over repeated indices. Plugging into $Q=\frac{1}{8\pi^2}\int_{\mathbb T^4} \mbox{tr}[F \wedge F]$, and using $\mbox{tr}[H_aH_b]=\delta_{ab}$ and $\nu_a\nu_a=1-1/N$, we find $Q=\frac{\bm k\cdot \bm m}{N}+\mathbb Z$.} \cite{tHooft:1981sps}:
\begin{eqnarray} \label{Q of n}
Q=\frac{1}{8\pi^2}\int_{\mathbb T^4} \mbox{tr}[F \wedge F]=-\frac{1}{8N}\epsilon_{\mu\nu\alpha\beta}n_{\mu\nu}n_{\alpha\beta}+\mathbb Z=\frac{\bm k\cdot \bm m}{N}+\mathbb Z\,,
\end{eqnarray}
where $F$ is the field strength of $A$. Notice that the twists $(\bm m, \bm k) \in (\mathbb Z\, \mbox{Mod}\, N)^6$. Adding multiples of $N$ to $\bm m$ or $\bm k$ leaves the cocycle condition intact. However, this has the effect of changing the topological charges by integers. Hence, from here on, we shall take the twists $m_i,k_i \in \mathbb Z$, not $\, \mbox{Mod}\, N$.  The partition function of the $SU(N)$ gauge theory with given twists $(\bm m, \bm k)$ is
\begin{eqnarray}
{\cal Z}[\bm m, \bm k]_{\scriptsize SU(N)}=\sum_{\nu\in \mathbb Z}\int \left[ D A_\mu\right]_{(\bm m, \bm k)} e^{-S_{YM}-i\left(\frac{\bm k\cdot \bm m}{N}+\nu\right)\theta}\,.
\end{eqnarray}
Here, $S_{YM}$ is the YM action, and the subscript $(\bm m, \bm k)$ indicates that the path integral is to be performed with a given set of twisted boundary conditions. Summation over the integer-valued topological sectors, $\nu \in \mathbb Z$, is necessary so that the theory satisfies locality (cluster decomposition). 

\subsubsection*{$SU(N)$ theory with matter}

Next, we add matter fields in a representation ${\cal R}$ under $SU(N)$. The matter representation has $N$-ality $n$. Then, the full $\mathbb Z_N$ center breaks down to $\mathbb Z_{q}$, $q=\mbox{gcd}(N,n)$, i.e., the Wilson lines are charged under $\mathbb Z_{q}^{(1)}$ $1$-from center symmetry\footnote{For example, $SU(2M)$ gauge theory with matter in the $2$-index (anti)symmetric representation has a $\mathbb Z_2^{(1)}$ center symmetry that acts on Wilson lines.}. Putting the matter, which, from now on, will be assumed to be fermions, on $\mathbb T^4$ modifies the cocycle conditions. Let $\psi$ be a left-handed Weyl fermion transforming under ${\cal R}$ of $SU(N)$. Then, the fermion obeys the boundary conditions
\begin{eqnarray}
\psi(x+\hat e_\mu L_\mu)={\cal R}(\Omega_{\mu}(x))\psi(x)\,.
\end{eqnarray}
The matrix ${\cal R}(\Omega_{\mu}(x))$ is built from $\Omega_\mu$, transforming in the defining representation of $SU(N)$, with suitable symmetrization or anti-symmetrization over $n$ indices (the $N$-ality of the representation) according to the specific representation ${\cal R}$. Thus, schematically (ignoring symmetrization over indices) 
\begin{eqnarray}
{\cal R}(\Omega_{\mu})\sim \underbrace{\Omega_{\mu}...\Omega_{\mu}}_{n}\,.
\end{eqnarray}
${\cal R}(\Omega_{\mu})$ must satisfy the cocycle condition
\begin{eqnarray}\label{cocycle general}
{\cal R}(\Omega_\mu (x + \hat{e}_\nu L_\nu)) \; {\cal R}(\Omega_\nu (x)) = {\cal R}(\Omega_\nu (x+ \hat{e}_\mu L_\mu)) \; {\cal R}(\Omega_\mu (x))\,,
\end{eqnarray}
 which, via Eq. (\ref{cocycle}), reveals that the allowed values of the twists $\bm m$ and $\bm k$ are $\frac{N}{q}, \frac{2N}{q}, ...$. Twisting by the center subgroup $\mathbb Z_q$ induces a background field with fractional topological charge
\begin{eqnarray}
Q=\frac{\bm m\cdot \bm k}{N}+\mathbb Z\,, \quad \bm m, \bm k\in \frac{N}{q}\mathbb Z\,,
\label{fractional Q}
\end{eqnarray}
and the partition function in the presence of matter reads
\begin{eqnarray}
\nonumber
{\cal Z}[\bm m, \bm k]_{\scriptsize SU(N)+\mbox{matter}}&=&\sum_{\nu\in \mathbb Z}\int \{\left[ D A_\mu\right] \left[ D\mbox{matter}\right]\}_{(\bm m, \bm k)} e^{-S_{YM}-S_{\scriptsize\mbox{matter}}-i\left(\frac{\bm k\cdot \bm m}{N}+\nu\right)\theta}\,,\\
&&  m_i,  k_i\in \frac{N}{q}\mathbb Z\,,\quad i=1,2,3\,.
\label{PF background m and k}
\end{eqnarray}

In the presence of matter,  the theory is endowed with classical nonabelian and abelian flavor symmetries. The $U(1)$ baryon-number symmetry survives the quantum corrections. In contrast, the chiral part of the abelian symmetry, denoted by $U(1)_\chi$, will generally break down to a discrete symmetry because of the Adler–Bell–Jackiw (ABJ) anomaly of $U(1)_\chi$ in the background of color instantons (which have integer topological charges). To fix ideas, we consider a single flavor of a Dirac fermion with classical $U(1)$ baryon number and $U(1)_\chi$ chiral symmetries. We take the $U(1)$ baryon charge of the Dirac fermion to be $+1$. The ABJ anomaly breaks $U(1)_\chi$ down to invertible $\mathbb Z_{2T_{\cal R}}^\chi$ chiral symmetry, where $T_{\cal R}$ is the Dynkin index of the representation. Generalizing the theory to include many flavors is straightforward, and we shall work out examples of this sort later in the paper. In the presence of the twists $(\bm m, \bm k)$, there can be an anomaly of $\mathbb Z_{2T_{\cal R}}^\chi$ in the background of $\mathbb Z_q^{(1)}$. The anomaly is a non-trivial phase acquired by ${\cal Z}[\bm m, \bm k]_{\scriptsize SU(N)+\mbox{matter}}$ as we apply a transformation by an element of  $\mathbb Z_{2T_{\cal R}}$:
\begin{eqnarray}
{\cal Z}[\bm m, \bm k]_{\scriptsize SU(N)+\mbox{matter}}|_{k_i,  m_i\in  N \mathbb Z/q}\longrightarrow e^{i 2\pi \ell \frac{ \bm m\cdot \bm k}{N}} {\cal Z}[\bm m, \bm k]_{\scriptsize SU(N)+\mbox{matter}}\,,
\end{eqnarray}
and $\ell=0,1,2,.., T_{\cal R}-1$ are the elements of $\mathbb Z_{2T_{\cal R}}^\chi$. For the smallest twists $m_j=k_j=\frac{N}{q}$ in the $j$-th direction, we obtain \cite{Anber:2021lzb}
\begin{eqnarray}
{\cal Z}[\bm m, \bm k]_{\scriptsize SU(N)+\mbox{matter}}|_{m_3=k_3=\frac{N}{q}}\longrightarrow e^{i 2\pi \ell \frac{N}{q^2}} {\cal Z}[\bm m, \bm k]_{\scriptsize SU(N)+\mbox{matter}}\,.
\label{the mixed anomaly from Z}
\end{eqnarray}
Bearing in mind that $N/q\in \mathbb Z$, we can generally absorb the integral part of $N/q^2$ by adding an integer topological charge, which cannot change the anomaly. Nevertheless, we will retain the phase as indicated in Eq. (\ref{the mixed anomaly from Z}).   The phase is nontrivial, and hence there is an anomaly, if and only if $\ell \frac{N}{q^2} \not\in \mathbb Z$.  In the next section, we show how to obtain the same anomaly using the Hamiltonian formalism.

We can do more regarding turning on fractional fluxes in $SU(N)$ with matter.  Instead of limiting ourselves to $\mathbb Z_q$ twists, we can twist with the full $\mathbb Z_N$ center symmetry or any subgroup of it provided we also turn on backgrounds of $U(1)$ baryon number symmetry \cite{Anber:2019nze,Anber:2021lzb}.   Let $\omega_\mu$ denote the $U(1)$ transition functions such that for the $U(1)$ gauge field $a_\mu$, we have $a_\nu(x+\hat e_\mu)=\omega_\mu\circ a_\nu(x)\equiv a_\nu(x)-i\omega_\mu^{-1}\partial_\nu\omega_\mu$. Then, $\Omega_\mu$ and $\omega_\mu$ obey the cocycle conditions:
\begin{eqnarray}\label{full cocycle}
\nonumber
\Omega_\mu (x + \hat{e}_\nu L_\nu) \; \Omega_\nu (x) = e^{i {2 \pi n_{\mu\nu} \over N}} \Omega_\nu (x+ \hat{e}_\mu L_\mu) \; \Omega_\mu (x)\,,\\
\omega_\mu (x + \hat{e}_\nu L_\nu) \; \omega_\nu (x) = e^{-i {2 \pi n n_{\mu\nu} \over N}} \omega_\nu (x+ \hat{e}_\mu L_\mu) \; \omega_\mu (x)\,,
\end{eqnarray}
where the $N$-ality of the matter representation is incorporated in the abelian transition functions.  The topological charges of both the  nonabelian center and abelian backgrounds read\footnote{Here, we use Footnote \ref{t hooft flux on T4} along with abelian field strengths $f_{12}=\frac{2\pi}{L_1L_2}(\frac{n}{N}m_3+A_3)$ and $f_{34}=\frac{2\pi}{L_3L_4}(\frac{n}{N}k_3+B_3)$ in the $1$-$2$ and $3$-$4$ planes, and similar expressions in the rest of the planes. Substituting into $ Q_u=\int_{\mathbb T^4}\frac{f\wedge f}{8\pi^2}$, we obtain the fractional $U(1)$ topological charge.}
\begin{eqnarray}\nonumber
Q_{SU(N)}=\frac{\bm m\cdot \bm k}{N}+\mathbb Z\,,\quad Q_u= \left(\frac{n}{N}\bm m+\bm A\right)\cdot \left(\frac{n}{N}\bm k+\bm B\right)\,,\quad \bm m, \bm k, \bm A,\bm B \in \mathbb Z^3\,.\\ \label{both nonabelian and u1 charges}
\end{eqnarray}
Here, $\bm A, \bm B$ are arbitrary integral magnetic and electric quantum numbers that we can always turn on since they leave the cocycle condition intact. 

\subsubsection*{$SU(N)\times U(1)$ theory with matter}

We may also choose to make $U(1)$ dynamical, which entails summing over small and large gauge transformations of $U(1)$, with the latter implementing integer winding. This results in $SU(N)\times U(1)$ gauge theory with a Dirac fermion in representation ${\cal R}$, with $N$-ality $N$ and baryon-charge $+1$. In this case, the $U(1)$ instantons reduce  $\mathbb Z_{2T_{\cal R}}^\chi$ down to the genuine (invertible) symmetry $\mathbb Z^\chi_{\scriptsize 2\mbox{gcd}(T_{\cal R},d_{\cal R})}$, and $d_{\cal R}$ is the dimension of $\cal R$. The easiest way to see that is by recalling the partition function under a $U(1)_\chi$ transformation acquires a phase:
 \begin{eqnarray}
 \exp\left[i2\alpha T_{\cal R} \int_{\mathbb T^4}\frac{\mbox{tr}\left(F\wedge F\right)}{8\pi^2}+i2\alpha d_{\cal R}  \int_{\mathbb T^4}\frac{f\wedge f}{8\pi^2} \right]\,,
 \end{eqnarray}
where $f$ is the field strength of the $U(1)$ field. 
Recalling that for the dynamical $SU(N)$ and $U(1)$ fields we have $ \int_{\mathbb T^4}\frac{\mbox{tr}\left(F\wedge F\right)}{8\pi^2}\in \mathbb Z, \int_{\mathbb T^4}\frac{f\wedge f}{8\pi^2} \in \mathbb Z$, we conclude that only  $\mathbb Z^\chi_{\scriptsize 2\mbox{gcd}(T_{\cal R},d_{\cal R})}$ survises the chiral transformation. The theory admits Wilson's lines:
\begin{eqnarray}
W_{\mu\,,SU(N)}=\mbox{tr}_\Box\left[P e^{i \int_{x_\mu=0}^{x_\mu=L_\mu} A_\mu}\Omega_{\mu}\right]\,,\quad W_{\mu\,,U(1)}=e^{-i \int_{x_\mu=0}^{x_\mu=L_\mu} a_\mu}\omega_{\mu}\,,
\label{Wilson lines for SUNU1}
\end{eqnarray}
which are charged under an electric  $\mathbb Z_N^{(1)}$ $1$-form center symmetry. In addition, the theory is endowed with a magnetic $U^{(1)}_m(1)$   $1$-form symmetry because of the absence of magnetic monopoles. For the sake of completeness, we also give the partition function of $SU(N)\times U(1)$ theory with matter in the background of given $(\bm m, \bm k)$ fluxes:
\begin{eqnarray}
\nonumber
{\cal Z}[\bm m, \bm k]_{\scriptsize SU(N)\times U(1)+\mbox{matter}}&=&\sum_{\nu, \nu_{U(1)}\in \mathbb Z}\int \{\left[ D A_\mu\right] \left[ D a_\mu\right] \left[ D\,\mbox{matter}\right]\}_{(\bm m, \bm k)} e^{-S_{YM}-S_{U(1)}-S_{\scriptsize\mbox{matter}}}\,,\\
&&  m_i,  k_i\in \mathbb Z\, \,,\quad i=1,2,3\,,
\label{PF background m and k with sunu1}
\end{eqnarray}
and in addition to the $SU(N)$ integer topological charges $\nu$, we included a sum over integer topological charges $\nu_{U(1)}$ of the $U(1)$ sector. 

\subsection{Twisting in the Hamiltonian formalism}
\label{The Hamiltonian formalism}

\subsubsection*{Pure $SU(N)$ theory}

Let us repeat the above discussion using the Hamiltonian formalism, starting with pure $SU(N)$ YM theory (we use a hat to distinguish an operator in this section.) To this end, we put the gauge theory on a spatial $3$-torus $\mathbb T^3$ and apply the magnetic $\bm m$ twists along the $3$-spatial directions. The transition functions in the defining representation along the spatial directions, denoted by $\Gamma_i$, can be chosen to be constant $N\times N$ matrices obeying the cocycle condition
\begin{eqnarray}\label{cocycle spacial}
\Gamma_i \; \Gamma_j = e^{i {2 \pi \epsilon_{ijk}m_k \over N}} \Gamma_j  \; \Gamma_i\,.
\end{eqnarray}
  Then, one can construct the states of the physical Hilbert space using the temporal gauge condition $A_0=0$.  The states can be written using the ``position" eigenstates of the gauge fields $A_j$, $j=1,2,3$ (or $i=x,y,z$) as follows:
\begin{eqnarray}\label{position eigenstates}
|\psi\rangle_{\bm m}\equiv |A_1, A_2, A_3\rangle_{\bm m}\,, \quad \hat A_j |A_1, A_2, A_3\rangle_{\bm m}=A_j|A_1, A_2, A_3\rangle_{\bm m}\,,
\end{eqnarray}
and the subscript $\bm m$ emphasizes that the Hilbert space is constructed in the background of the magnetic twists.  In writing Eq. (\ref{position eigenstates}), we have put many details under the rug, and the reader is referred to \cite{tHooft:1981sps,vanBaal:1984ra,Cox:2021vsa} for details. For example, notice that the gauge fields $A_i$  need to respect the twisted boundary conditions (\ref{cocycle spacial}), i.e., they transform according to (\ref{conditions on gauge field}) as we traverse any spatial direction on $\mathbb T^3$. The theory admits $3$ fundamental Wilson lines wrapping the three cycles of $\mathbb T^3$; these are given by (\ref{Wilson lines}) by restricting $\mu$ to the spatial directions.  The Wilson lines are charged under the $\mathbb Z_N^{(1)}$ $1$-form symmetry generated by three symmetry generators $\hat T_j$, the Gukov-Witten operators, supported on co-dimension $2$ surfaces. Thus, we have \begin{eqnarray}\label{comm of T and W}\hat T_j \hat W_j=e^{i \frac{2\pi}{N}} \hat W_j \hat T_j\,,\end{eqnarray} and there are $\hat W_j^{e_j}$ distinct Wilson's lines with $N$ distinct $N$-alities $e_j=0,1,..,N-1$. The center-symmetry generators $\hat T_i$ are hard to construct explicitly. However, their explicit form is not important to us. What is important is that they commute with the YM Hamiltonian $\hat H$, and thus, $\hat H$ and $\hat T_i$ can be simultaneously diagonalized. The physical states of the theory $|\psi\rangle_{\scriptsize \mbox{phy}\,,\bm m}$ are designated by the eigenvalues of $\hat T_i$. It can be shown that the action of $\hat T_i$ on $|\psi\rangle_{\scriptsize \mbox{phy}\,,\bm m}$ is given by
\begin{eqnarray}\label{action of T on physical states pure SUN}
\hat T_j |\psi\rangle_{\scriptsize \mbox{phy}\,,\bm m}=e^{i \frac{2\pi}{N}e_j-i\theta \frac{m_j}{N} } |\psi\rangle_{\scriptsize \mbox{phy}\,,\bm m}\,,
\end{eqnarray}
where $e_j, m_j \in \mathbb Z_N$  and the $\theta$ term ensures that $\hat T_i^N |\psi\rangle_{\scriptsize \mbox{phy}\,,\bm m}=e^{-i\theta m_j}|\psi\rangle_{\scriptsize \mbox{phy}\,,\bm m}$, and hence, $\hat T_i^N$ works as a large gauge transformation. The combination $e_j-\frac{\theta}{2\pi}m_j$ is the $\mathbb Z_N$ electric flux in the $j$-th direction. This is justified as follows. Consider the state $\hat W_j |\psi\rangle_{\scriptsize \mbox{phy}\,,\bm m}$, obtained from $ |\psi\rangle_{\scriptsize \mbox{phy}\,,\bm m}$ by the action of $\hat W_j$. Using Eqs. (\ref{comm of T and W}, \ref{action of T on physical states pure SUN}), we find $\hat T_j \hat W_j |\psi\rangle_{\scriptsize \mbox{phy}\,,\bm m}=e^{i \frac{2\pi}{N}(e_j+1)-i\theta \frac{m_j}{N} } \hat W_j|\psi\rangle_{\scriptsize \mbox{phy}\,,\bm m}$. Therefore,
acting with $\hat W_j$ on the state $ |\psi\rangle_{\scriptsize \mbox{phy}\,,\bm m}$ increases $e_j$ by one unit in the $j$-th direction. Since $\hat W_j$
inserts an electric flux tube winding in the $j$-th direction, the interpretation of $e_j$ as electric flux follows. Notice also that because $\hat T_j$ and $\hat H$ can be simultaneously diagonalized, we may label the states by the energy and the electric flux: \begin{eqnarray}\label{the physical psi for the first time}|\psi\rangle_{\scriptsize \mbox{phy}\,,\bm m}\equiv |E,\bm e\rangle_{\bm m}\,,\quad \bm e\in \mathbb Z_N^3\,.\end{eqnarray}
It is worth spending some time to explain our notation in Eq. (\ref{the physical psi for the first time}), as we shall use this notation extensively in our paper.  The physical state is labeled by the eigenvalues of a set of commuting operators, here the energy and the electric flux. The $SU(N)$ theory does not admit a $1$-form magnetic symmetry, and thus, we cannot label the states by magnetic fluxes. Yet, we can turn on a background magnetic flux $\bm m$, indicated as a subscript; all physical quantities are calculated in this magnetic background. Also, we use the letter $\bm m$ to denote the set of magnetic fluxes we can consistently turn on. Here, we have $\bm m\in \mathbb Z^3$.

 How can we make sense of the fractional topological charge (\ref{fractional Q}) on $\mathbb T^3$? We consider the product of $\mathbb T^3$ and the time interval $[0,L_4]$ and consider the boundary conditions $\hat A_i(t=L_4)=C[\bm k]\circ \hat A_i(t=0) $, where $C[\bm k]$ is an ``improper gauge" transformation implementing a twist $\bm k \in \mathbb Z^3$  on the gauge fields by an element of the center\footnote{In fact, $C$ should be designated by both $\bm k$ and the integral instanton number $\nu$; see \cite{tHooft:1981sps}. However, $\nu$ does not play a role in this work.}. In the presence of the magnetic twists $\bm m$, it can be shown that an application of $C[\bm k]$ results in the topological charge (Pontryagin square) \cite{tHooft:1981sps,vanBaal:1984ra,Cox:2021vsa}:
\begin{eqnarray}
Q[C[\bm k]]=\int_{\mathbb T^3} K(C\circ\hat A)-K(\hat A)= \frac{1}{24\pi^2}\int_{\mathbb T^3}\mbox{tr}\left[CdC^{-1}\right]^3=\frac{\bm m\cdot \bm k}{N}+\mathbb Z\,,
\label{main Q relation}
\end{eqnarray}
where $K(\hat A)$ is the topological current density operator $K(\hat A)=\frac{1}{8\pi^2}\mbox{tr}\left[\hat A\wedge \hat F-\frac{i}{3}\hat A\wedge \hat A\wedge \hat A\right]$, or in terms of the components: $\hat K^{\mu}(A)=\frac{1}{16\pi^2}\epsilon^{\mu\nu\lambda\sigma}\left(\hat A_\nu^a\partial_\lambda\hat A_\sigma^a-\frac{1}{3}f^{abc}\hat A_\nu^a \hat A_\lambda^b \hat A_\sigma^c\right)$.

\subsubsection*{$SU(N)$ theory with matter}

Adding fermions of $N$-ality $n$ changes the center from $\mathbb Z_N$ to $\mathbb Z_q$, $q=\mbox{gcd}(N,n)$, and the twists $(\bm m, \bm k)$ are now in $\left(N \mathbb Z/q\right)^6$. Otherwise, all the steps used to put the theory on $\mathbb T^3$ and construct the Hilbert space carry over. In particular, $\hat T_i$ now are the generators of the $\mathbb Z_q^{(1)}$ $1$-form symmetry, and their action on the physical states in the Hilbert space is given by\footnote{\label{footnote to ditinguish}It is conceivable to introduce an additional label to signify the distinct symmetries generated by different operators $\hat T_j$. For instance, we could designate $\hat T_{N, j}$ as the generator of $\mathbb Z_N^{(1)}$ and $\hat T_{q,j}$ as the generator of $\mathbb Z_q^{(1)}$. Nonetheless, this approach may lead to increased complexity in our expressions, and we opt not to pursue it. Instead, we will explicitly specify the symmetry in question when discussing these distinct operators.} (now we turn off the $\theta$ angle as we can rotate it away via a chiral transformation acting on the fermion)
\begin{eqnarray}
\hat T_j |\psi\rangle_{\scriptsize \mbox{phy}\,,\bm m}=e^{i \frac{2\pi}{q}e_j} |\psi\rangle_{\scriptsize \mbox{phy}\,,\bm m}\,,
\label{action of t in Zq on vectors}
\end{eqnarray}
and the theory has $e_j=0,1,2,..,q-1$ electric flux sectors in each direction $j=1,2,3$. The operators $\hat T_j$ act on the spatial Wilson lines in the defining representation of $SU(N)$ as $\hat T_j \hat W_j=e^{i \frac{2\pi}{q}} \hat W_j \hat T_j$, and there are $q$ distinct Wilson's lines $W_j^{e_j}$. The physical states $|\psi\rangle_{\scriptsize \mbox{phy}\,,\bm m}$ are simultaneous eigenstates of the Hamiltonian and $\hat T_j$ since both operators commute. Thus, we can write the physical states in the magnetic flux background $\bm m \in \left(N \mathbb Z/q\right)^3$ as
\begin{eqnarray}\label{eigenstates in sun with no twist}
|\psi\rangle_{\scriptsize \mbox{phy}\,,\bm m}=|E, \bm e N/q \rangle_{\bm m}\,, \quad \bm e=(e_1,e_2,e_3)\in \mathbb Z_q^3\,,
\end{eqnarray}
and $Ne_j/q$ is the amount of electric flux carried by the state in direction $j$. We may also say that $e_j$ is the number of electric fluxes in units of $N/q$. For matter with $N$-ality $n=0$, e.g., in the adjoint representation, $q=N$ and we recover what we have said about pure $SU(N)$ gauge theory.

The partition function (\ref{PF background m and k}) can be written in the Hamiltonian formalism as a trace over states in Hilbert space:
\begin{eqnarray}\nonumber
{\cal Z}[\bm m, \bm k]_{\scriptsize SU(N)+\mbox{matter}}&=&\mbox{tr}_{\bm m}\left[e^{-L_4 \hat H}(\hat T_x)^{k_x}(\hat T_y)^{k_y}(\hat T_z)^{k_z}\right]\\
&=&\sum_{\bm e\in \{0,1,.., q-1\}^3} e^{i\frac{2\pi \bm e\cdot \bm k}{q}}{}_{\bm m}\langle E, \bm e N/q |e^{-L_4 \hat H}|E, \bm e N/q \rangle_{\bm m}\,,
\end{eqnarray}
where the subscript $\bm m$ in the trace means that we are considering the states in the background of the magnetic flux $\bm m \in \left(N \mathbb Z/q\right)^3$. We also used Eqs. (\ref{action of t in Zq on vectors}, \ref{eigenstates in sun with no twist}), the fact that the states are eigenstates of both the energy and the $1$-form center operators.  

 To detect the anomaly between $\mathbb Z_{2T_{\cal R}}^\chi$ and $\mathbb Z_q^{(1)}$ in the Hamiltonian formalism, we first define the operator that implements the discrete chiral symmetry. To this end, we recall that under a chiral $U(1)_{A}$ rotation, the presence of the ABJ anomaly indicates non-conservation of the corresponding symmetry rotation:
\begin{equation}
\del_{\mu}\hat  j^{\mu}_{A} = 2T_{\cal R} \partial_{\mu}\hat  K^{\mu} (A)\,.
\end{equation}
Yet, we can define a conserved current:
\begin{eqnarray}
\hat j_5^\mu\equiv \hat j_A^\mu-2T_{\cal R}\hat K^\mu\,,
\end{eqnarray}  
and correspondingly a conserved charge:
\begin{eqnarray}
\hat Q_5=\int_{\mathbb T^3} \hat J_5^0\,.
\end{eqnarray}
Therefore, it is natural to define the operator
\begin{eqnarray}
\hat U_{\mathbb Z_{2T_{\cal R}}, \ell}\equiv \exp\left[i \frac{2\pi \ell}{2T_{\cal R}} \hat Q_5 \right]=  \exp\left[i \frac{2\pi \ell}{2T_{\cal R}} \int_{\mathbb T^3} (\hat  j_A^0-2T_{\cal R}\hat K^0(\hat A))  \right]\,,
\end{eqnarray}
for $\ell=0,1,.., T_{\cal R}-1$, which implements the action of the $\mathbb Z_{2 T_{\cal R}}^\chi$ chiral symmetry. $\hat U_{\mathbb Z_{2T_{\cal R}}}$ is invariant under both  small and large $SU(N)$ gauge transformations (with integer winding).  To find the mixed anomaly between $\mathbb Z_{2 T_{\cal R}}^\chi$  and $\mathbb Z_q^{(1)}$, we compute the commutation between $\hat T_j$, which implements the action of the electric center symmetry in the $j$-th direction, and $\hat U_{\mathbb Z_{2T_{\cal R}}}$:
\begin{eqnarray}
\hat T_j \hat U_{\mathbb Z_{2T_{\cal R}}, \ell}\hat T_j^{-1}\,,
\end{eqnarray}
remembering that the theory is in the background of a magnetic twist $m_j\in \frac{N\mathbb Z}{q}$ in the $j$-th direction\footnote{Similar to the Footnote \ref{footnote to ditinguish}, we could use a label that denotes the specific magnetic flux background when we are dealing with the operator $\hat U_{\mathbb Z_{2T_{\cal R}}, \ell}$. This background can be taken in sets such as $\frac{N\mathbb Z}{q}$ or $\frac{N\mathbb Z}{p}$, among others. However, adopting this approach may introduce unnecessary complexity to our notation. As a result, we have chosen to adopt a more transparent approach: we will explicitly mention the magnetic flux background whenever we discuss this operator.}. First, $\hat T_j$ commutes with the current $j_A^0$ since the latter is a color singlet operator. However, $\hat K^0$ fails to commute with $\hat T_j$; the commutation between the two operators is found by recalling that the action of $\hat T_j$ is implemented on the gauge fields $\hat A_{j}$ as $\hat A_j=C[k_j]\circ \hat A_j$. Thus, we find, after making use of (\ref{main Q relation}), 
\begin{eqnarray}
\nonumber
&&\hat T_j  \exp\left[i 2\pi \ell \int_{\mathbb T^3}   \hat K^0(\hat A) \right]\hat T_j^{-1}\\
\nonumber
&=&\exp\left[i 2\pi \ell \int_{\mathbb T^3}  \hat K^0\left(C[k_j]\circ \hat A\right)-\hat K^0(\hat A) \right]\exp\left[i 2\pi \ell \int_{\mathbb T^3}  \hat K^0(\hat A) \right]\\
&=&\exp\left[i 2\pi \ell \frac{m_j k_j}{N}\right]_{m_j, k_j \in \left(\frac{N \mathbb Z}{q}\right)^2}\exp\left[i 2\pi \ell \int_{\mathbb T^3}  \hat K^0(\hat A) \right]\,,
\label{the manipulations of U}
\end{eqnarray}
noting the restriction $m_j, k_j \in \left(\frac{N \mathbb Z}{q}\right)^2$ due to the presence of matter; otherwise, we would not satisfy the cocycle condition. Collecting everything and using the minimal twists $m_j=k_j=\frac{N}{q}$ we conclude
\begin{eqnarray}
\hat T_j \hat U_{\mathbb Z_{2T_{\cal R}},\ell}\hat T_j^{-1}=e^{i2\pi \ell\frac{N}{q^2}}  \hat U_{\mathbb Z_{2T_{\cal R}},\ell}\,,
\label{main T relation}
\end{eqnarray}
which is exactly the mixed anomaly between the $\mathbb Z_{2T_{\cal R}}^\chi$ chiral and the $\mathbb Z_q^{(1)}$ $1$-form center symmetries found in (\ref{the mixed anomaly from Z}) from the path integral formalism. The anomaly along with the commutation relations (remember that both $\mathbb Z_q^{(1)}$ and $\mathbb Z^\chi_{2T_{\cal R}}$ are good symmetries of the theory, and hence, the corresponding operators commute with the Hamiltonian)
\begin{eqnarray}
[\hat H, \hat T_j]=0\,, \quad [\hat H, \hat U_{\mathbb Z_{2T_{\cal R}}}]=0\,,
\end{eqnarray}
furnishes a finite-dimensional space with a minimum dimension of $q^2/\mbox{gcd}(q^2,N)$. This means that sectors in Hilbert space exhibit a $q^2/\mbox{gcd}(q^2,N)$-fold degeneracy. 

\subsubsection*{$SU(N)\times U(1)$ theory with matter}

Next, we discuss the Hamiltonian quantization of $SU(N)\times U(1)$ gauge theory with matter fields on $\mathbb T^3$ in the background of twists. In this case, we may twist with the full $\mathbb Z_N$ center symmetry provided we also turn on a background of $U(1)$. Thus, we replace the cocycle conditions (\ref{cocycle spacial}) with
\begin{eqnarray}\label{cocycle spacial total}
\nonumber
\Gamma_i \; \Gamma_j &=& e^{i {2 \pi \epsilon_{ijk}m_k \over N}} \Gamma_j  \; \Gamma_i\,,\\
\omega_i (x + \hat{e}_j L_j) \; \omega_j (x) &=& e^{-i {2 \pi n \epsilon_{ijk}m_k \over N}} \omega_j (x+ \hat{e}_i L_i) \; \omega_i (x)\,,
\end{eqnarray}
and we included the $N$-ality of the matter representation $n$ in the cocycle condition of the abelian field. This guarantees that the combined transition functions satisfy the correct cocycle conditions in the presence of matter.  Here, we can allow background center fluxes with $(\bm m, \bm k)\in \mathbb Z^6$ for all matter representations, thanks to the $U(1)$ gauge group.
We also introduce the operators $\hat T_{j}$ for $SU(N)$ and $\hat t_j$ for $U(1)$, $j=1,2,3$. The combinations $\hat T_j \hat t_j $ are the generators of the electric $\mathbb Z_{N}^{(1)}$ $1$-form global symmetry and act on the spatial Wilson lines in (\ref{Wilson lines for SUNU1}) as:  $\hat T_{j}W_{j, SU(N)}=e^{i \frac{2\pi}{N}}W_{j, SU(N)}\hat T_{j}$ and $\hat t_{j}W_{j, U(1)}=e^{-i \frac{2\pi}{N}}W_{j, U(1))}\hat t_{j}$. The action of $\hat t_j$ is implemented on the gauge fields, as usual, by improper gauge transformations of $\hat a_{j}$ as $\hat a_j=c[k_j]\circ \hat a_j$, and amounts to applying  $n k_j$ (Mod $N$) electric twists (notice the appearance of the $N$-ality). Unlike $\hat T_j$, the explicit form of $\hat t_j$ is simple:  \begin{eqnarray}\label{abelian lambda} \hat t_j\equiv e^{i\lambda_j(x)}\,, \quad \lambda_j(x)=\frac{-2\pi n}{N}\frac{x_j k_j}{L_j}\,.\end{eqnarray} Since $\mathbb Z_N^{(1)}$ is a good global symmetry, we can choose the states in Hilbert space to be eigenstates of the $\mathbb Z_N^{(1)}$ generators $\hat T_j\hat t_j$:
\begin{eqnarray}
\hat T_j\hat t_j|\psi\rangle_{\scriptsize\mbox{phy}\,,\bm m}=e^{i\frac{ 2\pi e_j}{N}} |\psi\rangle_{\scriptsize\mbox{phy}\,,\bm m}\,,
\end{eqnarray}
where $e_j=0,1,..., N-1$. Notice that the states are constructed in the ``fractional"  background magnetic flux $\bm m\in \mathbb Z^3$ (remember that in principle $m_i \in \mathbb Z$ Mod $N$, and thus, it implements the fractional magnetic twist. However, we can always add multiples of $N$ to $m_i$ without affecting the cocycle conditions, and hence, we drop the Mod $N$ restriction.) In addition, the theory has a magnetic $U(1)^{(1)}_m$ $1$-form global symmetry,  which can be used to characterize the physical states by an ``integer" value of the magnetic flux. Therefore, a state in the physical Hilbert space can be labeled as \begin{eqnarray}|\psi\rangle_{\scriptsize\mbox{phy}\,,\bm m}=|E, \bm e, \bm N \rangle_{\bm m}\,, \quad \bm e \in \mathbb Z_N^3\,,\end{eqnarray} and $\bm N=(N_x, N_y, N_z) \in \mathbb Z^3$ (not Mod $N$) label the integral magnetic fluxes of the $U(1)$ gauge group.  The partition function (\ref{PF background m and k with sunu1}) can be written as a trace over states in Hilbert space in $(\bm m, \bm k)$ backgrounds as follows: 
\begin{eqnarray}\nonumber
{\cal Z}[\bm m, \bm k]_{\scriptsize SU(N)\times U(1)+\mbox{matter}}&=&\mbox{tr}_{\bm m}\left[e^{-L_4 \hat H}(\hat T_x\hat t_x)^{k_x}(\hat T_y\hat t_y)^{k_y}(\hat T_z\hat t_z)^{k_z}\right]\\\nonumber
&=&\sum_{\bm e\in \{0,1,.., N-1\}^3, \bm N\in \mathbb Z^3} e^{i\frac{2\pi \bm e\cdot \bm k}{N}}{}_{\bm m}\langle E, \bm e, \bm N |e^{-L_4 \hat H}|E, \bm e, \bm N \rangle_{\bm m}\,.\\
\label{the partition function in SUNXU1}
\end{eqnarray}

We also build the operator that corresponds to the chiral transformation. This construction was detailed in \cite{Anber:2023pny}, and we do not repeat it here. Instead, we only give a synopsis of the derivation, which is needed in this work.  The anomaly equation of the chiral current is
\begin{eqnarray}
\partial_\mu \hat  j^\mu_A-2T_{\cal R}\partial_\mu \hat K^\mu(\hat A)-\frac{2d_{\cal R}}{8\pi^2}\epsilon_{\mu\nu\lambda\sigma}\partial^\mu \hat a^\nu\partial^\lambda \hat a^\sigma=0\,.
\end{eqnarray}
Then, the chiral symmetry operator in the background of the $m_j$ magnetic flux is given by
\begin{eqnarray}
\hat U_{\mathbb Z_{2T_{\cal R}}, \ell} =\exp\left[i \frac{2\pi \ell}{2T_{\cal R}} \hat Q_5 \right]\,,
\end{eqnarray}
where the conserved charge $\hat Q_5$ is given by
\begin{eqnarray}\label{q5}\nonumber
\hat Q_5 &=& \int_{\mathbb T^3} d^3 x \left[ \hat j_\chi^0 - 2 T_{\cal R} K^{0}(\hat A) - 2 \frac{d_{\cal R}}{8\pi^2}\epsilon^{ijk} \hat a_i \partial_j \hat a_k \right]\\
&&+  { d_R \over 4 \pi} (N_{z} + {n \over N} n_{z})\left[ \int\limits_{0}^{L_y} {d y \over L_y} \int\limits_{0}^{L_z} d z \hat a_z(x=0,y,z) +\int\limits_{0}^{L_x} {d x \over L_x} \int\limits_{0}^{L_z} d z \hat a_z(x,y=0,z)  \right]  \nonumber\\
&&+ \sum\limits_{\scriptsize\mbox{cyclic}}  (x \rightarrow y \rightarrow z \rightarrow x)\,.
\end{eqnarray}
The last term comes from carefully treating the boundary term implied from the transition functions $\omega_j(x)$, since, unlike $\Gamma_j$,  they depend explicitly on $x_j$, see \cite{Anber:2023pny} for details. In addition to the background flux $n_j$, which introduces the fractional winding number, we also allow integer magnetic winding $N_j$. Under a transformation with $\hat t_j$, the integral of the abelian Chern-Simons term $\hat K^0(\hat a)=\epsilon^{ijk} \hat a_i \partial_j \hat a_k$ in the background of the integral $M_j$ and fractional $m_j$ magnetic fluxes  transforms as (recall (\ref{abelian lambda}))
\begin{eqnarray}\label{the fractional abelian charge}
\nonumber
\hat t_j \exp\left[i\int_{\mathbb T^3}\hat K^0(\hat a)\right] t_j^{-1}&=&\exp\left[i\int_{\mathbb T^3}\hat K^0(c\circ \hat a)-i\int_{\mathbb T^3}\hat K^0(\hat a)\right]\exp\left[i \int_{\mathbb T^3}\hat K^0(\hat a) \right]\\
&&=\left( N_j+\frac{n  n_j}{N} \right) \left(\frac{n k_j}{N}\right)\exp\left[i\int_{\mathbb T^3}\hat K^0(\hat a) \right]\,.
\end{eqnarray}

The reader will notice that we switched from the letter $\bm m$, which we use to signify the set of fractional fluxes we can activate, e.g., here we have $\bm m \in \mathbb Z^3$, to the letter $\bm n$, which is the actual number of fractional magnetic fluxes we turn on. We shall use the same labeling throughout the paper.

In the next sections, we use these constructions to argue that $SU(N)/\mathbb Z_p$, $\mathbb Z_p\subseteq \mathbb Z_q$ as well as $SU(N)\times U(1)/\mathbb Z_p$, $\mathbb Z_p\subseteq \mathbb Z_N$ enjoy a noninvertible $0$-form chiral symmetry, with a possible mixed anomaly with the $1$-form center symmetry. 

\section{$SU(N)/\mathbb Z_p$, $\mathbb Z_p\subseteq \mathbb Z_q$ theories, noninvertible symmetries, and their anomalies}
\label{SUNMATHBB ZP THEORIES}

In this section, we direct our attention to YM theories featuring matter fields residing in a particular representation ${\cal R}$ and characterized by an $N$-ality  $n$. Building upon the discussion in the preceding section, it is established that $SU(N)$ gauge theories, when coupled to matter, exhibit an electric $\mathbb Z_q^{(1)}$ $1$-form center symmetry (recall $q=\mbox{gcd}(N,n)$). A notable maneuver within this framework involves the gauging of $\mathbb Z_q^{(1)}$ or a subgroup of it, leading to  $SU(N)/\mathbb Z_p$ theory, $\mathbb Z_p\subseteq \mathbb Z_q$, whose partition function is obtained by summing over integer and fractional topological charge sectors.
Thus, gauge transformations with fractional winding numbers are part of the gauge structure, and well-defined operators should be invariant under such gauge transformations. Here, we would like to emphasize that there are $p$ distinct theories: $\left(SU(N)/\mathbb Z_p\right)_{n}$, $n=0,1,...,p$, which differ by the admissible genuine (electric, magnetic, or dyonic)  line operators. In this paper, we limit our treatment to $\left(SU(N)/\mathbb Z_p\right)_{n=0}$, and whenever we mention $SU(N)/\mathbb Z_p$, we particularly mean $\left(SU(N)/\mathbb Z_p\right)_{0}$. 
What happens to the invertible $\mathbb Z_{2 T_{\cal R}}^\chi$ discrete chiral symmetry of this theory? As we shall discuss, this symmetry can stay invertible or become noninvertible, depending on whether it exhibits a mixed anomaly with $\mathbb Z_p^{(1)}$ symmetry in the original $SU(N)$ theory.

\subsection{$SU(N)/\mathbb Z_q$}

We start by discussing noninvertible $0$-form chiral symmetries in $SU(N)/\mathbb Z_q$ theories, i.e., theories obtained by gauging the full electric $\mathbb Z_q^{(1)}$ $1$-form center symmetry. Such theories do not possess global electric $1$-form symmetry; hence, there are no genuine Wilson's lines. This can be understood as follows. We start with pure $SU(N)$ gauge theory, which has an electric $\mathbb Z_N^{(1)}$ $1$-form symmetry and admits the full spectrum of Wilson lines, i.e., it admits Wilson's lines with all $N$-alities $n=0,1,2,..,N-1$. Gauging a $\mathbb Z_q$ subgroup of $\mathbb Z_N$, we obtain $SU(N)/\mathbb Z_q$ gauge theory. Now, the spectrum of allowed Wilson lines must be invariant under $\mathbb Z_q$, forcing us to remove those lines with $N$-alities that are not multiples of $q$. The remaining lines in pure $SU(N)/\mathbb Z_q$ theory are charged under an electric $\mathbb Z_{N/q}^{(1)}$ $1$-form symmetry; these are $W_j^{qe_j}$, with $e_j=0,1,..,N/q-1$ and $W_j$ is Wilson's line in the defining representation of $SU(N)$. Finally, introducing matter with $N$-ality $q$ means that those remaining lines can end on the matter and must also be removed from the spectrum. This deprives $SU(N)/\mathbb Z_q$ gauge theory with matter from all genuine Wilson's lines.

Despite that $SU(N)/\mathbb Z_q$ theory with matter does not possess an electric $1$-form symmetry, it is endowed with a magnetic $\mathbb Z_q^{m(1)}$ $1$-form global symmetry. This can be understood, again, starting from the pure $SU(N)/\mathbb Z_q$ theory. As we discussed above, the pure theory has an electric $\mathbb Z_{N/q}^{(1)}$ $1$-form symmetry. The magnetic dual of $SU(N)/\mathbb Z_q$ is $SU(N)/\mathbb Z_{N/q}$, which admits a magnetic $\mathbb Z_q^{m(1)}$ $1$-form symmetry. The pure $SU(N)/\mathbb Z_q$ theory has $q$ distinct magnetic fluxes ('t Hooft lines) in its spectrum. Let ${\cal T}_j$ be the 't Hooft line winding around direction $j$ in the defining representation of $SU(N)$, i.e., it has $N$-ality $1$. Then, the pure $SU(N)/\mathbb Z_q$ theory possesses the following set of 't Hooft lines  ${\cal T}_j^{n_jN/q}$, $n_j=0,1,.., q-1$ for $j=1,2,3$, which are mutually local with the set of Wilson's lines $W_j^{qe_j}$, $e_j=0,1,..,N/q-1$\footnote{This can be easily seen since ${\cal T}_j^{n_jN/q}$ and $W_j^{qe_j}$ satisfy the Dirac quantization condition.}. Introducing electric matter removes all Wilson's lines (as stated above) but does not alter the magnetic symmetry. Thus, we conclude that $SU(N)/\mathbb Z_q$ theory with matter possesses a magnetic $\mathbb Z_q^{m(1)}$ $1$-form global symmetry acting on a set of 't Hooft lines ${\cal T}_j^{n_jN/q}$, $n_j=0,1,.., q-1$ for $j=1,2,3$.     

We can label the states in the physical Hilbert space of $SU(N)/\mathbb Z_q$ theory with matter by both energy and magnetic fluxes since the Hamiltonian commutes with the generators of the magnetic $\mathbb Z_q^{m(1)}$ $1$-form symmetry\footnote{Recall that the allowed magnetic twists in the $SU(N)$ theory with matter are $\bm m\in (N\mathbb Z/q)^3$.}:
\begin{eqnarray}\label{physical states labeled with magnetic lines}
| \psi\rangle_{\scriptsize \mbox{phy}}=|E,\bm n N/q\rangle\,,\quad \bm n=(n_x,n_y,n_z) \in (\mathbb Z_q)^3\,.
\end{eqnarray}
The partition function of these theories involves summing over sectors with fractional topological charges $N\mathbb Z/q^2$ (use Eq.(\ref{fractional Q}) and set $k_i=m_i=N/q$), which can be written in the path-integral formalism as (we set the vacuum angle $\theta=0$) 
\begin{eqnarray}\label{PF of sunmodzq}
\nonumber
{\cal Z}_{\scriptsize SU(N)/\mathbb Z_q+\mbox{matter}}&=&\sum_{\nu\in \mathbb Z, (\bm m, \bm k)\in \left( N\mathbb Z/q\right)^6}\int \{\left[ D A_\mu\right] D\left[\mbox{matter}\right]\}_{(\bm m, \bm k)} e^{-S_{YM}-S_{\scriptsize\mbox{matter}}}\,,\\
\end{eqnarray}
or in the Hamiltonian formalism as
\begin{eqnarray}\label{PF hamiltonian sunzq}
{\cal Z}_{\scriptsize SU(N)/\mathbb Z_q+\mbox{matter}}=\mbox{tr}\left[e^{-L_4 \hat H}\right]=\sum_{\scriptsize \mbox{physical states}} {}_{\scriptsize \mbox{phy}}\langle\psi |e^{-L_4 \hat H}|\psi\rangle_{\scriptsize \mbox{phy}}\,.
\end{eqnarray}
%
%
%

Our main task is to build a gauge invariant operator that implements the $\mathbb Z^\chi_{2T_{\cal R}}$ chiral transformation in $SU(N)/\mathbb Z_q$ theory with matter. 
To this end, we use the Hamiltonian formalism of Section \ref{The Hamiltonian formalism}, dropping the hats from all operators to reduce clutter. We also use $x,y,z$ to label the three spatial directions. For $\ell \in \Z_{2T_{\rep}}^\chi$, the chiral symmetry operator is given by:
\begin{equation}
U_{\Z_{2T_{\rep}},\ell} = e^{2\pi i \f{\ell}{2T_{\rep}} \int_{\mathbb T^3} \left(j^{0}_{A} - 2T_{\rep} K^{0}(A) \right)}\,.
\end{equation}
This operator is invariant under large gauge transformations with integer winding numbers. 
We will now gauge the $\Z_{q}^{(1)}$ one-form symmetry. In $SU(N)/\Z_{q}$ gauge theory with matter, we sum over arbitrary $\mathbb Z_q$ twists with fractional topological charges $N\mathbb Z/q^2$. We consider the operator $U_{\Z_{2T_{\rep}},\ell}$ in the presence of magnetic fluxes $\bm m\in \left(N\mathbb Z/q\right)^3$ (these are the magnetic fluxes that label the physical states in Eq. (\ref{physical states labeled with magnetic lines}).)  Let $T_{x}$ be the generator of an electric $\mathbb Z_q$ center twist along the $x$ direction (i.e., a $\mathbb Z_{q}$ gauge transformation), and we take it to have the minimal twist of $N/q$. It acts on $U_{\Z_{2T_{\rep}},\ell} $ via (recall the discussion around Eq. (\ref{the manipulations of U}))
\begin{equation}
T_{x} U_{\Z_{2T_{\rep}},\ell}   T_{x}^{-1} = e^{-2\pi i \ell Q} U_{\Z_{2T_{\rep}},\ell}\ = e^{-2\pi i \ell \f{n_{x} N}{q^2}} U_{\Z_{2T_{\rep}},\ell}\,,\quad  n_x\in \mathbb Z\,. 
\label{the main t Hooft anomaly relation in x}
\end{equation}
$n_x$ counts the magnetic fluxes inserted in the $y$-$z$ plane in units of $N/q$.
Identical relations to (\ref{the main t Hooft anomaly relation in x}) hold in the $y$ and $z$ directions. 
As we saw in the previous section, if $\ell \f{N}{q^2} \not \in \mathbb Z$,  there is a mixed 't Hooft anomaly between the electric $\Z_{q}^{(1)}$ $1$-form center and the discrete chiral symmetries of $SU(N)$ theory with matter. Eq. (\ref{the main t Hooft anomaly relation in x}) implies that the operator $U_{\Z_{2T_{\rep}},\ell} $ is not gauge invariant under a $\mathbb Z_q$ gauge transformation as we attempt to gauge $\Z_{q}^{(1)}$.  We can remedy this problem and reconstruct a gauge-invariant operator, denoted by $\tilde U_{\Z_{2T_{\rep}}}$,  by summing over all $\Z_{q}$ gauge transformations generated by $T_{x}$, $T_{y}$ and $T_{z}$:
\begin{align}
    \tilde U_{\Z_{2T_{\rep}},\ell} & \equiv \sum_{p_{x}, p_{y},p_{z} \in \Z} (T_{x})^{p_{x}} (T_{y})^{p_{y}} (T_{z})^{p_{z}} U_{\Z_{2T_{\rep}}}  (T_{x})^{-p_{x}} (T_{y})^{-p_{y}} (T_{z})^{-p_{z}} \nonumber \\
    & = U_{\Z_{2T_{\rep}},\ell} \sum_{p_{x}, p_{y},p_{z}\in \mathbb Z}  e^{-2\pi i \frac{\ell N}{q^2} \left(p_x n_x+p_yn_y+p_zn_z\right)}\equiv U_{\Z_{2T_{\rep}},\ell}\sum_{\bm p\in \mathbb Z^3}  e^{-2\pi i \frac{\ell N}{q^2} \bm p\cdot\bm n} \nonumber\\
    & = U_{\Z_{2T_{\rep}},\ell} \sum_{l_{x} \in \Z} \d \left( \f{n_{x}\ell N}{q^2} - l_{x} \right) \sum_{l_{y} \in \Z} \d \left( \f{n_{y}\ell N}{q^2} - l_{y} \right) \sum_{l_{z} \in \Z} \d \left( \f{n_{z}\ell N}{q^2} - l_{z} \right)\,.
 \label{def of U one of main results}
\end{align}
In the first line, we included a sum over arbitrary powers of $T_x, T_y,T_z$ to enforce the gauge invariance.  
Then, we used Eq. (\ref{the main t Hooft anomaly relation in x}) in going from the first to the second line and the Poisson resummation formula in going from the second to the third line. Even though $\tilde U_{\Z_{2T_{\rep}},\ell}$ is gauge invariant, it has no inverse; it is, in general, a noninvertible operator that implements the action of $\tilde{\Z}_{2T_{\rep}}^\chi$, and we use a tilde to denote the noninvertible nature of symmetries and their operators. The noninvertibility stems from the fact that $\tilde U_{\Z_{2T_{\rep}}}$ works as a projector: the insertion of this operator in the path integral of $SU(N)/\mathbb Z_q$ theory with matter projects onto specific topological charge sectors of $SU(N)/\mathbb Z_q$, depending on $\ell$. This can be seen from the second line in (\ref{def of U one of main results}), which is a sum over Fourier modes that projects in and out sectors, depending on their topological charge, upon acting on them. One can see the projective nature of  $\tilde U_{\Z_{2T_{\rep}},\ell}$ by inserting it into the partition function (\ref{PF hamiltonian sunzq}):
\begin{eqnarray}
\langle    \tilde U_{\Z_{2T_{\rep}},\ell}  \rangle=\sum_{\scriptsize \mbox{physical states}} {}_{\scriptsize \mbox{phy}}\langle\psi |e^{-L_4 \hat H}  \tilde U_{\Z_{2T_{\rep}},\ell}  |\psi\rangle_{\scriptsize \mbox{phy}}\,,
\end{eqnarray}
and then using the physical states defined in Eq. (\ref{physical states labeled with magnetic lines}). We find that $ \tilde U_{\Z_{2T_{\rep}},\ell}$ annihilates sectors with $ \f{n_{x}\ell N}{q^2} \notin \mathbb Z$, etc. We remind that $\f{n_{x}N}{q^2}$ is the topological charge (see Eq. (\ref{fractional Q})), which we can write as
\begin{eqnarray}
\f{n_{x} N}{q^2}=\underbrace{\frac{n_xN}{q}}_{m_x}\underbrace{\frac{N}{q}}_{k_x}\frac{1}{N}\,,
\end{eqnarray}
and, as we mentioned earlier and emphasize now, $n_x$ is the number of magnetic fluxes in units of $N/q$. The same applies to the magnetic sectors in the $y$ and $z$ directions. We conclude that $  \tilde U_{\Z_{2T_{\rep}},\ell} $ selects sectors in Hilbert space with certain magnetic fluxes.

  We can make the following observations about $\tilde U_{\Z_{2T_{\rep}},\ell}$:
\begin{enumerate}
    \item  If $\ell \in q\Z$, $\tilde U_{\Z_{2T_{\rep}},\ell}$ is invertible since in this case $\f{n_{x,y,z}\ell N}{q^2} \in \Z$ for all values of $n_{x}, n_y,n_z\in \mathbb Z$. 
     The invertible subgroup of $\tilde{\Z}_{2T_{\rep}}^\chi$ is $\Z_{2T_{\rep}/q}^\chi$.
    \item If $\gcd (\ell N/q, q) = 1$, then we must have $n_{x}, n_y,n_z \in q\Z$. In other words, $\tilde U_{\Z_{2T_{\rep}},\ell}$ projects onto untwisted flux sectors. In particular,  in the sector given by $n_{x}, n_y,n_z \in q\Z$, the symmetry operator $ \tilde U_{\Z_{2T_{\rep}},\ell} $ act invertibly for all elements of the chiral symmetry $\ell=1,2,.., T_{\cal R}$.
  \item If $\gcd (\ell N/q, q) = a \neq 1$ and $\ell<q$, then let $q = aq'$, and we must have $n_{x,y,z} \in q' \Z$. $\tilde U_{\Z_{2T_{\rep}},\ell}$ projects onto background fluxes with topological charge $Q \in \Z/q'$, i.e. sectors that have $\mathbb Z_{q'}$ twists.   
  \item The noninvertibility of $\tilde U_{\Z_{2T_{\rep}},\ell}$ can be seen by multiplying the operator by its inverse to find
  \begin{eqnarray}
\tilde U_{\Z_{2T_{\rep}},\ell}\times \tilde U_{\Z_{2T_{\rep}},\ell}^{-1}\sim\sum_{\bm p\in \mathbb Z^3}  e^{-2\pi i \frac{\ell N}{q^2} \bm p\cdot\bm n} \equiv {\cal C}\,.
  \end{eqnarray}
 ${\cal C}$ is known as the condensation operator, which can be thought of as a sum over topological surface operators $\exp[-i \oint_{\mathbb T^2\subset \mathbb T^3} B^{(2)}]=\exp[-i2\pi\mathbb Z/q]$ wrapping the three $2$-cycles of $\mathbb T^3$, and $B^{(2)}$ is the $2$-form field of the $\mathbb Z^{(1)}_q$ $1$-form symmetry. 
\end{enumerate}

We use the fact that $SU(N)/\mathbb Z_q$ theory possesses a magnetic $\mathbb Z_q^{m(1)}$ $1$-form global symmetry to make one more observation. Let ${\cal T}_{j}$ be 't Hooft line of $N$-ality $1$ in direction $j$. Then, the minimal 't Hooft line in $SU(N)/\mathbb Z_q$ theory is ${\cal T}_{j}^{N/q}$, i.e., it has $N$-ality $N/q$. The minimal line acts on a physical state by increasing its magnetic flux by one in units of $N/q$\footnote{Similar to the discussion we had after Eq. (\ref{action of T on physical states pure SUN}), we can also consider the generators of the magnetic $1$-form symmetry and argue that ${\cal T}_j^{N/q}$ inserts a magnetic flux $N/q$, as measured by the action of the magnetic $1$-form symmetry on the state ${\cal T}_j^{N/q}|\psi\rangle_{\scriptsize \mbox{phy}}$.}. Now, let us take a theory with $\mbox{gcd}(N/q,q)=1$ so that $\tilde U_{\Z_{2T_{\rep}},\ell=1}$ acts projectively on certain states. Then,  $|E,(n_x=q,n_y=q,n_z=q)N/q \rangle$ is one of the physical states that survive under the action of  $\tilde U_{\Z_{2T_{\rep}},\ell=1}$. We have ${\cal T}_x^{N/q}|E,(q,q,q)N/q\rangle=|E,(q+1,q,q)N/q\rangle$. Thus, we immediately see from Eq. (\ref {def of U one of main results}) that \begin{equation}\tilde U_{\Z_{2T_{\rep}},\ell=1}{\cal T}_x^{N/q}|E,(q,q,q)N/q\rangle=\tilde U_{\Z_{2T_{\rep}},\ell=1}|E,(q+1,q,q)N/q\rangle\rangle=0\,.\end{equation} We write this result as
 \begin{eqnarray}
 \tilde U_{\Z_{2T_{\rep}},\ell=1}{\cal T}_{j}^{N/q}=0\,,\quad j=x,y,z\,.
 \end{eqnarray}
 In other words, the operator $ \tilde U_{\Z_{2T_{\rep}},\ell=1}$ annihilates the minimal 't Hooft lines in this theory. It also annihilates all 't Hooft lines ${\cal T}_{j}^{n_jN/q}$, $n_j\neq 0$ Mod $q$.   This is an alternative way to see the projective nature of this operator.

\subsection{$SU(N)/\mathbb Z_p$}

Next, we discuss $SU(N)/\mathbb Z_p$ theory with matter with $N$-ality $n$, and $\mathbb Z_p\subseteq \mathbb Z_q=\mathbb Z_{\scriptsize\mbox{gcd}(N,n)}$. The partition function of this theory is given by the path integral in Eq. (\ref{PF of sunmodzq}), now restricting the sum over the electric and magnetic twists $(\bm m, \bm k) \in (N\mathbb Z/p)^6$.  The theory possess an electric $\mathbb Z_{q/p}^{(1)}$ $1$-form global symmetry. As before, $T_{x}$ is taken to be the generator of the electric $\mathbb Z_q^{(1)}$ symmetry. Then, the electric $\mathbb Z_{q/p}^{(1)}$ $1$-form global symmetry is generated by $T_x^{p}$ (as well as $T_y^{p}$ and $T_z^{p}$). 
The theory has $q/p$ distinct Wilson's lines $W_j^{e_j p}$, with $e_j=0,1,2,..,q/p-1$. These lines are invariant under $\mathbb Z_p$, as they should be since $\mathbb Z_p$ is gauged. The minimal admissible Wilson's line $W_j^p$ carries one electric flux in units of $pN/q$.  In the limiting case $p=q$, the line  $W_j^{p=q}$ coincides with the matter content and must be removed from the spectrum of line operators. Therefore, in this case, the theory does not possess a $1$-form electric symmetry, as discussed in the previous section.  

In addition, the theory has a magnetic $\mathbb Z_p^{m(1)}$ $1$-form symmetry. If ${\cal T}_j$ is the 't Hooft line with $N$-ality $1$, then the minimal admissible 't Hooft line in the theory is ${\cal T}_j^{N/p}$, which carries one magnetic flux in units of $N/p$. There are $p$ distinct 't Hooft lines in the theory ${\cal T}_j^{n_jN/p}$, $n_j=0,1,.., p-1$, which are mutually local with Wilson's lines  $W_j^{e_j p}$. The Hamiltonian, Wilson's lines generators, and the 't Hooft lines generators of this theory can be simultaneously diagonalized. Therefore,  the energies and eigenvalues of the set of Wilson and 't Hooft operators can be used to label the physical states of Hilbert space:
\begin{eqnarray}\label{psiphysical in sunmodp}
|\psi\rangle_{\scriptsize\mbox{phy}}=|E,  \bm e pN/q, \bm n N/p\rangle\,, \quad  \bm e \in (\mathbb Z_{q/p})^3\,, \bm n \in (\mathbb Z_p)^3\,.
\end{eqnarray}

 Next, we need to build a gauge invariant chiral symmetry operator. Our starting point, as usual, is the operator
\begin{equation}
U_{\Z_{2T_{\rep}},\ell} = e^{2\pi i \f{\ell}{2T_{\rep}} \int_{\mathbb T^3} \left(j^{0}_{A} - 2T_{\rep} K^{0}(A) \right)}
\end{equation}
 taken in the presence of the fractional magnetic fluxes $\bm m\in \left(N\mathbb Z/p\right)^3$, which label the Hilbert space in Eq. (\ref{psiphysical in sunmodp}). The operator $T_x^{q/p}$ generates the electric $\mathbb Z_p^{(1)}$ $1$-form symmetry, which is gauged. In other words, $T_x^{q/p}$ implements the twists $\bm k\in \left(N\mathbb Z/p\right)^3$.  In analogy with $SU(N)/\mathbb Z_q$ theories, we need to build gauge invariants of the chiral symmetry operator using the building block $T_{x}^{q/p} U_{\Z_{2T_{\rep}},\ell}   T_{x}^{-q/p}$. To compute this block, we use the discussion around Eq. (\ref{the manipulations of U}), taking the minimal twist $N/p$ generated by $T_x^{q/p}$,  to obtain
\begin{equation}
T_{x}^{q/p} U_{\Z_{2T_{\rep}},\ell}   T_{x}^{-q/p} =  e^{-2\pi i \ell \f{n_{x} N }{p^2}} U_{\Z_{2T_{\rep}},\ell}\,,\quad n_x\in \mathbb Z\,,
\label{the main t Hooft anomaly relation in x for zp}
\end{equation}
and $n_x$ counts the magnetic fluxes in units of $N/p$.
If $ \ell \f{ N}{p^2}\not\in\mathbb Z$, there is a mixed anomaly between $\mathbb Z_{2T_R}^\chi$ and the electric $\mathbb Z_p^{(1)}$ symmetries in $SU(N)$ theory with matter, and we expect the chiral symmetry becomes noninvertible upon gauging $\mathbb Z_p^{(1)}$. The corresponding gauge invariant operator of the $\tilde{\Z}_{2T_{\rep}}^\chi$ symmetry is then given by the summations
\begin{align}
    \tilde U_{\Z_{2T_{\rep}},\ell} &= \sum_{p_{x}, p_{y},p_{z} \in \Z} (T_{x})^{qp_{x}/p} (T_{y})^{qp_{y}/p} (T_{z})^{qp_{z}/p} U_{\Z_{2T_{\rep}},\ell}  (T_{x})^{-qp_{x}/p} (T_{y})^{-qp_{y}/p} (T_{z})^{-qp_{z}/p} \nonumber \\\nonumber
    & = U_{\Z_{2T_{\rep}},\ell} \sum_{p_{x}, p_{y},p_{z}\in \mathbb Z}  e^{-2\pi i \frac{\ell N}{p^2} \left(p_x n_x+p_yn_y+p_zn_z\right)}\\
    & = U_{\Z_{2T_{\rep}},\ell} \sum_{l_{x} \in \Z} \d \left( \f{n_{x}\ell N}{p^2} - l_{x} \right) \sum_{l_{y} \in \Z} \d \left( \f{n_{y}\ell N}{p^2} - l_{y} \right) \sum_{l_{z} \in \Z} \d \left( \f{n_{z}\ell N}{p^2} - l_{z} \right)\,.
 \label{def of U one of main results for zp}
\end{align}
This noninvertible operator generalizes (\ref{def of U one of main results}) to any $\mathbb Z_p\subseteq \mathbb Z_q$, and it projects onto sectors with finer topological charges than the sectors admissible by (\ref{def of U one of main results}). This means there exist sectors where $\tilde U_{\Z_{2T_{\rep}},\ell}$ act invertibly for all $\ell=1,2,.., T_{\cal R}$ if and only if
\begin{eqnarray}
l_x=\frac{n_{x} N}{p^2}\in \mathbb Z\,,
\label{condition for integer l in umodzp}
\end{eqnarray}
with similar conditions in the $y$ and $z$ directions. 
  As special cases, we may first set $p=q$ to readily cover (\ref{def of U one of main results}). Also, setting $p=1$, the operator $\tilde U_{\Z_{2T_{\rep}},\ell}$ becomes invertible, as can be easily seen from the second line in (\ref{def of U one of main results for zp}). Notice that $\tilde U_{\Z_{2T_{\rep}},\ell} $ does not act on Wilson's lines in this theory, as the noninvertible operator is built from $ (T_{j})^{qp_{j}/p}$ and its inverse; thus, one can push a Wilson line through $\tilde U_{\Z_{2T_{\rep}},\ell} $ without being affected\footnote{Although we do not give the explicit form of $T_j$, it can be thought of as an exponential of an integral of the chromoelectric field over a $2$-dimensional submanifold; see \cite{Reinhardt:2002mb}. A Wilson line would acquire a phase as we push it past $T_j^{q/p}$ (we use $[A^a_j(\bm x, t), E^b_k(\bm y,t)]=i\delta_{jk}\delta(\bm x-\bm y)\delta_{ab}$, where $a,b$ are the color indices, along with the Baker-Campbell-Hausdorff formula). It also acquires the negative of the same phase as it is pushed past $T_j^{-q/p}$. Therefore, the phases cancel out, and hence, the result in Eq. (\ref{U commute with W}).}. We can write this observation as
 \begin{eqnarray}\label{U commute with W}
 \tilde U_{\Z_{2T_{\rep}},\ell} W_j^{e_j p}=   W_j^{e_j p}\tilde U_{\Z_{2T_{\rep}},\ell}\,,\quad e_j=0,1,2,..,q/p-1\,, \quad j=x,y,z\,.
 \end{eqnarray}
  This is very different from the action of $\tilde U_{\Z_{2T_{\rep}},\ell} $ on 't Hooft lines, as we discussed before.

The procedure employed to construct the noninvertible operator $\tilde U_{\Z_{2T_{\rep}},\ell}$ contains an additional layer of underlying physics. It is essential to keep in mind that this operator is constructed in $SU(N)/\mathbb Z_p$ theory, where its creation involved a sum over magnetic $\bm m \in \left(N\mathbb Z/p\right)^3$ and electric $\bm k \in \left(N\mathbb Z/p\right)^3$ twists. These twists do not encompass the entire range of permissible twists that can be applied. Recall that the theory encompasses a global $\mathbb Z_{q/p}^{(1)}$ symmetry, which affords us the opportunity to introduce the electric twists $\bm k \in \left( pN\mathbb Z/q\right)^3$. Moreover, we can turn on magnetic twists $\bm m \in  \left( pN\mathbb Z/q\right)^3$, compatible with the cocycle condition\footnote{Recall that twists in $N\mathbb Z/q$ are compatible with the cocycle conditions. Therefore, twists in $ p N \mathbb Z/q$ are a subset of the allowed twists. Notice that the twists $\bm m \in (p N\mathbb Z/q)^3$ provide background magnetic fluxes and do not label the physical states in Hilbert space, Eq. (\ref{psiphysical in sunmodp}).}. This broader scope of twists provides a richer set of possibilities within the theory. We recall that $T_x^p$ is the generator of $\mathbb Z_{q/p}^{(1)}$ symmetry that implements the twists $k_x\in   pN\mathbb Z/q$. Then, one can write the partition function of $SU(N)/\mathbb Z_p$ theory in these background twists as
\begin{eqnarray}
\nonumber
{\cal Z}_{\scriptsize SU(N)/\mathbb Z_p+\mbox{matter}}[\bm m, \bm k]&=&\mbox{tr}_{\bm m \in \left(pN\mathbb Z/q\right)^3}\left[e^{-L_4 H}T_x^{k_x p}T_y^{k_y p}T_z^{k_z p}\right]\\
&=&\sum_{\bm e  \in (\mathbb Z_{q/p})^3}e^{-i2\pi \frac{p\bm k\cdot \bm e}{q}}{}_{\scriptsize\mbox{phy}}\langle \psi |e^{-L_4 H}|\rangle_{\scriptsize\mbox{phy}}|_{\bm m \in \left(pN\mathbb Z/q\right)^3}\,,
\end{eqnarray}
and we used Eq. (\ref{psiphysical in sunmodp}) along with $T_j^{k_j p}|\psi\rangle_{\scriptsize\mbox{phy}}=e^{-i2\pi \frac{p k_j e_j}{q}}|\psi\rangle_{\scriptsize\mbox{phy}}$; see the discussion around Eqs. (\ref{comm T and H}, \ref{action of tp in Zq on vectors}) below.

Next, consider the commutation relation between $T_x^{p}$ and $ \tilde U_{\Z_{2T_{\rep}},\ell}$, the latter operator is being in the background of the magnetic twist $\bm m\in  \left( pN\mathbb Z/q\right)^3$. Using the discussion and procedure around Eq. (\ref{the manipulations of U}), we obtain 
\begin{eqnarray}
T_x^{p} \tilde U_{\Z_{2T_{\rep}},\ell}T_x^{-p}= e^{-2\pi i \ell n_x \f{p^2N}{q^2}} \tilde U_{\Z_{2T_{\rep}},\ell}\,.
\label{the mixed anomaly zp theories}
\end{eqnarray}
The failure of the commutation between $T_x^{p}$ and $ \tilde U_{\Z_{2T_{\rep}},\ell}$ by the phase $ e^{-2\pi i \ell n_x \f{p^2N}{q^2}}$, assuming $\ell n_x \f{p^2N}{q^2}\notin \mathbb Z$ , signals a mixed anomaly between the noninvertible  $\tilde{\Z}_{2T_{\rep}}^\chi$ chiral symmetry and the electric $\mathbb Z_{q/p}^{(1)}$ $1$-form global symmetry. This anomaly means that certain sectors in Hilbert space exhibit degeneracy. 
Let us analyze this situation more closely. We assume there exists a sector with $n_{x},n{_y}, n_{z}$ that satisfies Eq. (\ref{condition for integer l in umodzp}), and thus, in this sector, the symmetry operator $\tilde U_{\Z_{2T_{\rep}},\ell}$ acts invertibly for all elements $\ell=1,2,.., T_{\cal R}$. Now, $\tilde U_{\Z_{2T_{\rep}},\ell}$, being a global symmetry operator, commutes with the Hamiltonian: 
\begin{eqnarray}
[\tilde U_{\Z_{2T_{\rep}},\ell}, H]=0\,.
\label{comm H U}
\end{eqnarray}
Likewise, since $\mathbb Z_{q/p}^{(1)}$ is a  global symmetry, its generators $T_j^{p}$ commute with the Hamiltonian: 
\begin{eqnarray}
[T_j^{p}, H]=0\,.
\label{comm T and H}
\end{eqnarray}
This commutation relation, along with Eq. (\ref{action of t in Zq on vectors}), implies that $T_j^p$ acts on physical states in Hilbert space as (the label $\bm l=(l_x,l_y,l_z)$ emphasizes that such states satisfy condition (\ref{condition for integer l in umodzp}), such that $\tilde U_{\Z_{2T_{\rep}},\ell}$ acts invertibly on such states. Also, we suppressed the detailed dependence on the different quantum numbers to reduce clutter)
\begin{eqnarray}
T_j^p | E, e_j\rangle_{\bm l}=e^{i \frac{2\pi p}{q}e_j} |E, e_j\rangle_{\bm l}\,,
\label{action of tp in Zq on vectors}
\end{eqnarray}
and that the states are labeled by their energies as well as $e_j=1,2,...,q/p$ distinct labels; these are the eigenvalues (fluxes) of the $\mathbb Z_{q/p}^{(1)}$ symmetry operator. 
The algebra defined by the commutation relations Eqs. (\ref{comm H U}, \ref{comm T and H}), along with the mixed anomaly represented as Eq. (\ref{the mixed anomaly zp theories}), under the assumption of a nontrivial phase, furnishes a finite-dimensional space with a minimum dimension of $q^2/\mbox{gcd}(n_xp^2N,q^2)$ (we take $n_x=n_y=n_z$). The Hilbert space of physical states, which are labeled by $q/p$ distinct fluxes, sit in $q^2/\mbox{gcd}(n_xp^2N,q^2)$ orbits, and a rotation by $\tilde U_{\Z_{2T_{\rep}},\ell=1}$ links a state with a flux $e_j$ to a state with a flux $e_{\scriptsize j+\mbox{gcd}(n_jp^2N,q^2)/(qp)}$ as:
\begin{eqnarray}
\tilde U_{\Z_{2T_{\rep}},\ell=1} |E, e_j\rangle_{\bm l}= |E, e_{\scriptsize j}+\mbox{gcd}(n_jp^2N,q^2)/(qp)\rangle_{\bm l}\,.
\end{eqnarray}
Using the commutation relation (\ref{comm H U}), one immediately sees that the states $|E, e_j\rangle_{\bm l}$ and $ |E, e_{\scriptsize j}+\mbox{gcd}(n_jp^2N,q^2)/(qp)\rangle_{\bm l}$ have the same energy\footnote{It is helpful to give a numerical example. Take $N=1000$, $q=500$, and $p=20$. Such numbers are contrived and do not necessarily correspond to any realistic theory. Condition (\ref{condition for integer l in umodzp}) is satisfied if we take $n_x=2$. Then, the phase in the anomaly Eq. (\ref{the mixed anomaly zp theories}) is $e^{-i2\pi/5}$, implying a $5$-fold degeneracy. The theory has an electric $\mathbb Z^{(1)}_{25}$ $1$-form symmetry, and thus, $25$ distinct flux states. These states set in $5$ different orbits such that the states labeled with $e_1$, $e_6$, $e_{11}$, $e_{16}$, $e_{21}$ have the same energy, and the states $e_2$,$e_7$,...,$e_{22}$, have the same energy, etc.}. 

In the following subsections, we apply our formalism to examples of theories with fermions in specific representations.

\subsection{Examples}

\subsubsection{$SU(4n+2)/\mathbb Z_2$ and $SU(4n)/\mathbb Z_2$ with a Dirac fermion in the $2$-index anti-symmetric representation}
\label{YM theories with fermions in the 2index antisymm}

The $SU(4n+2)/\mathbb Z_2$ gauge theory with a $2$-index anti-symmetric Dirac fermion ($N$-ality $2$) has a $\Z_{8n}^\chi$ chiral symmetry. The $SU(4n+2)$ theory possesses an electric $\Z_{2}^{(1)}$ one-form symmetry. In \cite{Argurio:2023lwl}, the authors argued that upon gauging $\Z_{2}^{(1)}$, the odd rotations of $\Z_{8n}^\chi$ become non-invertible. We can show this is the case on $\mathbb T^3$ using our construction. Setting $N=4n+2$ in (\ref{def of U one of main results}), we obtain
\begin{equation}
    \tilde{U}_{\mathbb Z_{8n}, \ell} = U_{\mathbb Z_{8n}, \ell}  \sum_{l_{x} \in \Z} \d \left( \f{n_{x}\ell }{2} - l_{x} \right) \sum_{l_{y} \in \Z} \d \left( \f{n_{y}\ell }{2} - l_{y} \right) \sum_{l_{z} \in \Z} \d \left( \f{n_{z}\ell}{2} - l_{z} \right)\,.
\end{equation}
For $\ell$ odd, $\tilde{U}_{\mathbb Z_{8n}, \ell}$  projects onto untwisted gauge sectors and becomes non-invertible. 

The $SU(4n)/\mathbb Z_2$ theory with a $2$-index anti-symmetric Dirac fermion has a $\Z_{8n-4}^\chi$ chiral symmetry. The cocycle conditions, say in the $x$-direction, must satisfy (see Eq. (\ref{cocycle general}))
\begin{equation}
    e^{2\pi i \f{2n_{yz}}{4n}} = 1\,.
\end{equation}
Therefore we must have $n_{yz} \in 2n \Z$. There is no mixed anomaly between $\Z_{8n-4}^\chi$ and the electric $\Z_{2}^{(1)}$ symmetries in the $SU(4n)$ theory since the anomaly phase $\f{n_{yz}}{2} \in n\Z$ is trivial. Thus,  the full chiral symmetry $\mathbb Z_{8n-4}^\chi$ is invertible. This is also in agreement with \cite{Argurio:2023lwl}.

\subsubsection{$SU(6)/\mathbb Z_3$  with a Dirac fermion in the $3$-index anti-symmetric representation}

This theory has a  $\mathbb Z_6^\chi$ chiral symmetry. What is special about this theory is that its bilinear fermion operator vanishes identically because of Fermi statistics. Moreover, the $SU(6)$ theory exhibits a mixed anomaly between its electric $\mathbb Z_3^{(1)}$ $1$-form center and chiral symmetries \cite{Yamaguchi:2018xse,Anber:2019nfu}. Assuming confinement, then the chiral symmetry must be broken in the infrared. Yet, this breaking has to be accomplished via higher-order condensate.  Using (\ref{def of U one of main results}), we find that the operator corresponding to a chiral transformation in $SU(6)/\mathbb Z_3$ theory is
\begin{equation}
    \tilde{U}_{\mathbb Z_{6}, \ell} = U_{\mathbb Z_{6}, \ell}  \sum_{l_{x} \in \Z} \d \left( \f{2n_{x}\ell }{3} - l_{x} \right) \sum_{l_{y} \in \Z} \d \left( \f{2n_{y}\ell }{3} - l_{y} \right) \sum_{l_{z} \in \Z} \d \left( \f{2n_{z}\ell}{3} - l_{z} \right)\,.
\end{equation}
Hence, for $\ell\in \{1,2,4,5\}$, the operator $\tilde{U}_{\mathbb Z_{6}, \ell}$  projects onto untwisted gauge sectors, and the chiral symmetry operator becomes noninvertible. 

\subsubsection{$2$-index $SU(6)$ chiral gauge theory}

Our next example is a chiral gauge theory. This is $SU(6)$ YM theory with a single left-handed Weyl fermion $\psi$ in the $2$-index symmetric representation and  $5$ flavors of left-handed Weyl fermions $\chi$ in the complex conjugate $2$-index anti-symmetric representation. The fermion budget ensures the theory is free from gauge anomalies. The theory encompasses continuous global symmetry $SU(5)_\chi\times U(1)_A$, where $SU(5)_\chi$ acts on $\chi$. The charges  of $\psi$ and $\chi$ under $U(1)_A$ are $
q_\psi=-5\,, q_\chi=2$. The theory is also endowed with a $\mathbb Z_4^{\chi}$ chiral symmetry, which is taken to act on $\chi$ with a unit charge. It can be checked that this is a genuine symmetry since neither $\mathbb Z_4$ nor a subgroup of it can be absorbed in rotations in the centers of $SU(6)\times SU(5)_\chi$. It turns out, see \cite{Anber:2023yuh} for details (also see \cite{Anber:2021iip}), that we must divide the global symmetry by $\mathbb Z_3\times \mathbb Z_5$ to remove redundancies. Putting everything together and remembering that the theory possesses an electric $\mathbb Z_2^{(1)}$ $1$-form center symmetry (since all fermions have $N$-ality $n=2$), we write the faithful global group as:
\begin{eqnarray}
G^{\scriptsize\mbox{g}}=\frac{SU(5)_\chi\times U(1)_A}{\mathbb Z_3 \times \mathbb Z_5}\times \mathbb Z_4^{\chi}\times \mathbb Z_2^{(1)}\,.
\label{G global for SU56 k2}
\end{eqnarray}
This theory has an anomaly between its $\mathbb Z_2^{(1)}$ center symmetry and $\mathbb Z_4^{\chi}$ chiral symmetry. To see the anomaly, we recall that we can turn on the magnetic and electric twists $(\bm m, \bm k)\in (3\mathbb Z)^6$. This gives the topological charge $Q\in \mathbb Z/2$. Thus, under a chiral transformation, the partition function acquires a phase 
\begin{eqnarray}
{\cal Z}[\bm m, \bm k]\longrightarrow \exp\left[i \frac{2\pi \ell N_\chi T_\chi Q}{4}\right] {\cal Z}[\bm m, \bm k]=\exp\left[i 2\pi\ell/2 \right] {\cal Z}[\bm m, \bm k]\,,
\end{eqnarray}
where $N_\chi=5$ is the number of the $\chi$ flavors and $T_{\chi}=4$ is the Dynkin index of $\chi$. Therefore, we expect that  $\mathbb Z_4^{\chi}$ becomes noninvertible in the $SU(6)/\mathbb Z_2$ chiral theory.  Using (\ref{def of U one of main results}), the noninvertible operator corresponding to a chiral transformation in $SU(6)/\mathbb Z_2$ theory is
\begin{equation}
    \tilde{U}_{\mathbb Z_{4}, \ell} = U_{\mathbb Z_{4}, \ell}  \sum_{l_{x} \in \Z} \d \left( \f{n_{x}\ell }{2} - l_{x} \right) \sum_{l_{y} \in \Z} \d \left( \f{n_{y}\ell }{2} - l_{y} \right) \sum_{l_{z} \in \Z} \d \left( \f{n_{z}\ell}{2} - l_{z} \right)\,.
\end{equation}
Hence, for $\ell\in \{1,3\}$, the operator $\tilde{U}_{\mathbb Z_{4}, \ell}$  projects onto untwisted gauge sectors, and the chiral symmetry operator becomes noninvertible. 

\section{$SU(N)\times U(1)/\mathbb Z_p$, $\mathbb Z_p\subseteq \mathbb Z_N$ theories, noninvertible symmetries, and their anomalies}
\label{SUNTIMES U1MODZP}

In this section, we also gauge the $U(1)$ baryon number symmetry. Thus, we are discussing $SU(N)\times U(1)$ gauge theory with a Dirac fermion in a representation $\cal R$, $N$-ality $n$, and $U(1)$ charge $+1$. This theory, as we discussed in Section \ref{Preleminaries}, is endowed with an invertible $\mathbb Z^\chi_{\scriptsize 2\mbox{gcd}(T_{\cal R},d_{\cal R})}$ chiral symmetry as well as an electric $\mathbb Z_N^{(1)}$ center symmetry acting on its Wilson's lines; see Eqs. (\ref{Wilson lines for SUNU1}). However, in \cite{Anber:2023pny}, it was shown that $SU(N)\times U(1)$ theories also have noninvertible $\tilde{\mathbb Z}^\chi_{2T_{\cal R}}$ chiral symmetry. In the following, we first review the construction of the noninvertible $\tilde{\mathbb Z}^\chi_{2T_{\cal R}}$  operator in $SU(N)\times U(1)$ theories, and next, we discuss this operator in $SU(N)\times U(1)/\mathbb Z_p$, $\mathbb Z_p\subseteq \mathbb Z_N$, theories. 

\subsection{$SU(N)\times U(1)$}

Our starting point is the $SU(N)\times U(1)$ theory and its $\mathbb Z^\chi_{2T_{\cal R}}$ operator $U_{\mathbb Z_{2T_{\cal R}}, \ell} =e^{i \frac{2\pi \ell}{2T_{\cal R}}  Q_5 }$, where $Q_5 $ is the conserved chiral charge defined in Eq. (\ref{q5}) in the background of the fractional $n_{x,y,z}$ and integer $N_{x,y,z}$ magnetic fluxes in the $x,y,z$ directions. We remind that we can turn on fractional fluxes in $\mathbb Z_N$ irrespective of the $N$-ality of the matter content since we use $U(1)$ transition functions to impose the cocycle condition; see Eq. (\ref{cocycle spacial total}).  No nontrivial electric twists are applied at this stage, i.e., we take $\bm k \in (N \mathbb Z)^3$, since our nonabelian gauge group is $SU(N)$ rather than $SU(N)/\mathbb Z_p$. The operator $U_{\mathbb Z_{2T_{\cal R}}, \ell}$ is invariant under $SU(N)$. To see that, we apply a large $SU(N)$ gauge transformation, recalling Eq. (\ref{main Q relation}) and setting $\bm k \in (N \mathbb Z)^3$, which immediately gives the change in the nonabelian winding number by $Q\in \mathbb Z$. In addition, $U_{\mathbb Z_{2T_{\cal R}}, \ell} $  must be invariant under $U(1)$ gauge symmetry. The photon gauge field $a_i$ transforms under $U(1)$ gauge symmetry as $a_j(x+\hat e_k L_k)=a_j- \partial_k \xi(x)$, and $\xi(x)$ is a periodic gauge function: $\xi(x+\hat e_k L_k)=\xi(x)+2\pi p$, $ p \in \mathbb Z$. Applying a large $U(1)$ gauge transformation to $Q_5$, we find (see \cite{Anber:2023pny} for the derivation)
\begin{eqnarray}\label{failure under U1 trans}
U_{\mathbb Z_{2T_{\cal R}}, \ell}\longrightarrow U_{\mathbb Z_{2T_{\cal R}}, \ell}e^{-2\pi i\ell\left( p_x \f{d_{\rep}}{T_{\rep}} \left( N_{x} + \f{nn_{x}}{N}\right)+ p_y \f{d_{\rep}}{T_{\rep}} \left( N_{y} + \f{nn_{y}}{N}\right) + p_z \f{d_{\rep}}{T_{\rep}} \left( N_{z} + \f{nn_{z}}{N} \right)\right)}\,,
\end{eqnarray}
where $p_{x,y,z}$ are arbitrary integers corresponding to the $U(1)$ gauge transformation. Eq. (\ref{failure under U1 trans}) shows that the operator $U_{\mathbb Z_{2T_{\cal R}}, \ell}$ fails to be gauge invariant under $U(1)$ gauge symmetry. To remedy this problem, we follow the procedure of the previous section and define a new operator $\tilde U_{\mathbb Z_{2T_{\cal R}}, \ell}$ by summing arbitrary copies of the gauge-transformed $U_{\mathbb Z_{2T_{\cal R}}, \ell}$:
 \begin{eqnarray}
 \nonumber
 \tilde U_{\mathbb Z_{2T_{\cal R}}, \ell}&=& U_{\mathbb Z_{2T_{\cal R}}, \ell}\sum_{p_x,p_y,p_z\in \mathbb Z} e^{-2\pi i\ell\left( p_x \f{d_{\rep}}{T_{\rep}} \left( N_{x} + \f{nn_{x}}{N}\right)+ p_y \f{d_{\rep}}{T_{\rep}} \left( N_{y} + \f{nn_{y}}{N}\right) + p_z \f{d_{\rep}}{T_{\rep}} \left( N_{z} + \f{nn_{z}}{N} \right)\right)}\\
 &=&U_{\mathbb Z_{2T_{\cal R}}, \ell}\sum_{l_x\in \mathbb Z}\delta\left(\ell\f{d_{\rep}}{T_{\rep}} \left( N_{x} + \f{nn_{x}}{N}\right)-l_x\right)\left(\sum_{l_y\in \mathbb Z}...\right)\left(\sum_{l_z\in \mathbb Z}...\right)\,.
 \label{U tilde in SUNU1}
 \end{eqnarray}
 The operator $ \tilde U_{\mathbb Z_{2T_{\cal R}}, \ell}$ implements the chiral transformation of the now-noninvertible $\tilde{\mathbb Z}^\chi_{2T_{\cal R}}$ symmetry, as it acts projectively by selecting certain nonvanishing sectors in Hilbert space labeled by the integers $l_{x,y,z}$, such that for $\ell=1$ we must have
 \begin{eqnarray}\label{condition on lx}
l_x=\f{d_{\rep}}{T_{\rep}} \left( N_{x} + \f{nn_{x}}{N}\right)\in \mathbb Z\,,
 \end{eqnarray}
 with identical expressions for $l_y$ and $l_z$. Condition (\ref{condition on lx}) ensures that all the symmetry elements $\ell=1,2,.., T_{\cal R}$ act invertibly on the same admissible sector.   To explicitly see the projective nature of $\tilde U_{\mathbb Z_{2T_{\cal R}}, \ell}$ on states in Hilbert space, we use the partition function of the $SU(N)\times U(1)$ theory given by Eq. (\ref{the partition function in SUNXU1}) (we set the electric flux background $\bm k$=0 and, as usual, we use $\bm n$ to label a specific fractional magnetic flux background: $\bm n=(n_x,n_y,n_z)$) to compute $\langle\tilde U_{\mathbb Z_{2T_{\cal R}}, \ell}\rangle$:\footnote{Recall from our earlier analysis that the theory is endowed with electric $\mathbb Z_N^{(1)}$ and magnetic $U^{(1)}_m(1)$ symmetries, and the states of the theory are labeled by the eigenstates of these symmetries, $\bm e$ and $\bm N$, respectively.}
 \begin{eqnarray}
\langle\tilde U_{\mathbb Z_{2T_{\cal R}}, \ell}\rangle= \sum_{\bm e\in\mathbb Z_N^3,\, \bm N\in \mathbb Z^3} {}_{\bm n}\langle E, \bm e, \bm N |e^{-L_4 \hat H} \tilde U_{\mathbb Z_{2T_{\cal R}}, \ell}|E, \bm e, \bm N \rangle_{\bm n}\,.
 \end{eqnarray}  
 We immediately see from the Kronecker deltas in Eq. (\ref{U tilde in SUNU1}) that only those sectors with $\bm N$ satisfying Eq. (\ref{condition on lx}) are selected. 
 
  Turning off the fractional magnetic flux background (i.e., setting $\bm n=0$), the operator $\tilde U_{\mathbb Z_{2T_{\cal R}}, \ell}$ becomes invertible for $\ell \in T_{\cal R} \mathbb Z/\mbox{gcd}(T_{\cal R}, d_{\cal R})$. We recognize that  we have just recovered the invertible $\mathbb Z_{\scriptsize 2 \mbox{gcd}(T_{\cal R}, d_{\cal R})}^\chi$ subgroup of $\tilde{\mathbb Z}^\chi_{2T_{\cal R}}$. Furthermore, setting  $\bm n=0$, the operator $\tilde U_{\mathbb Z_{2T_{\cal R}}, \ell=1}$ destructs all Hilbert space sectors characterized with integral magnetic fluxes $\bm N \notin T_{\cal R}  \mathbb Z^3/\mbox{gcd}(T_{\cal R}, d_{\cal R})$. This noninvertible nature of the chiral operator should have been anticipated. When we start with the SU(N) theory with matter, we find an 't Hooft anomaly between its invertible $\mathbb Z_{2T_{\cal R}}^\chi$ chiral symmetry and $U(1)$ baryon symmetry. This anomaly is valued in $\mathbb Z_{\scriptsize T_{\cal R}/\mbox{gcd}(T_{\cal R}, d_{\cal R})}$. Upon gauging $U(1)$, this anomaly becomes of the ABJ type, and the chiral symmetry becomes noninvertible.  Now, If we take the Euclidean version of our theory in the infinite volume limit and apply a $\pi/2$ rotation to $\tilde U_{\mathbb Z_{2T_{\cal R}}, \ell=1}$, the operator becomes a defect. Alternatively, we may also use the half-gauging procedure to construct this defect, which was done in \cite{Anber:2023pny}.  Inserting this defect at some position will generally create a domain wall (since it enforces a chiral transformation) dressed with a TQFT that accounts for the noninvertible nature of the defect. It will be interesting to analyze what happens to the domain walls when we turn on an external magnetic field with flux $\bm N \notin  T_{\cal R}  \mathbb Z^3/\mbox{gcd}(T_{\cal R}, d_{\cal R})$.

 $SU(N)\times U(1)$ gauge theory has an electric $\mathbb Z_N^{(1)}$ $1$-form global center symmetry, and the immediate exercise would be checking whether there is a mixed anomaly between the center and the noninvertible chiral symmetries. To this end, we turn on both electric and magnetic twists\footnote{Notice that these electric twists $\bm k \in \mathbb Z^3$ are $\bm e$ that label the physical states in Hilbert space: $|E, \bm e, \bm N \rangle_{\bm n}$\,. In principle, $k_j$ should be in $\mathbb Z$ Mod $N$, but, as usual, we drop the modding as this does not affect the cocycle conditions.} $(\bm m, \bm k)\in\mathbb {\mathbb Z}^6$, giving rise to nonabelian fractional topological charge $Q_{SU(N)}\in \mathbb Z/N$ as well as abelian topological charge $Q_u=\left(\frac{n}{N}\right)^2$; see Eq. (\ref{both nonabelian and u1 charges}). Using Eqs. (\ref{main Q relation}, \ref{the fractional abelian charge}), setting $\bm k=(1,0,0)$, we find
\begin{eqnarray}
\nonumber
T_{x}t_{x} \tilde U_{\mathbb Z_{2T_{\cal R}}, \ell} (T_{x}t_{x})^{-1} &=& \tilde U_{\mathbb Z_{2T_{\cal R}}, \ell} e^{-2\pi i \ell \left( \f{n_{x}}{N} - \f{n}{N} \f{d_{\rep}}{T_{\rep}} \left(N_x + \f{nn_{x}}{N}\right) \right)}\\
&=&  \tilde U_{\mathbb Z_{2T_{\cal R}}, \ell}e^{-i2\pi\ell\left(\frac{n_x-nl_x}{N} \right)}\,,
\label{sunu1 anomaly}
\end{eqnarray}
and we used Condition (\ref{condition on lx}) to go from the first to the second line. If the phase is nontrivial, then there is a mixed anomaly between the electric $\mathbb Z_N^{(1)}$ $1$-form center and the $0$-form noninvertible $\tilde{\mathbb Z}^\chi_{2T_{\cal R}}$  symmetries, leading to spectral degeneracy of states (those that already selected by the operator $\tilde U_{\mathbb Z_{2T_{\cal R}}, \ell}$). The algebra defined by the commutation relations $[H, T_jt_j]=[H,\tilde U_{\mathbb Z_{2T_{\cal R}}, \ell} ]=0$ along with the mixed anomaly (\ref{sunu1 anomaly}), under the assumption of a nontrivial phase, furnishes a finite-dimensional space with dimension $N/\mbox{gcd}(N,n_x-n l_x)$ (we take $n_x=n_y=n_z$). The Hilbert space of physical states, which are labeled by $N$ different electric fluxes $\bm e$, sit in $N/\mbox{gcd}(N,n_x-nl_x)$ orbits, and a rotation by $\tilde U_{\Z_{2T_{\rep}},\ell=1}$ links a state with a flux $e_j$ to a state with a flux $e_{\scriptsize j}+\mbox{gcd}(N,n_j-n l_j)$, i.e., they have the same energy.

\subsection{$SU(N)\times U(1)/\mathbb Z_p$, $\mathbb Z_p\subseteq \mathbb Z_N$}

Next, we study the noninvertible operators in $SU(N)\times U(1)/\mathbb Z_p$ gauge theory, where $\mathbb Z_p$ is a subgroup of the $\mathbb Z_N$ center symmetry. This theory has an electric $\mathbb Z_{N/p}^{(1)}$ $1$-form global symmetry acting on the $p$-th power of the spatial components of the abelian and nonabelian Wilson's lines defined in Eq. (\ref{Wilson lines for SUNU1}):
\begin{eqnarray}\label{abelian and nonabelian lines in sunu1modzp}
W_{j,\,SU(N)}^{e_j p}\,, W_{j,\,U(1)}^{e_j p}\,,\quad e_j=1,2,..,N/p,\quad j=1,2,3\,.
\end{eqnarray}
These Wilson's lines are invariant under $\mathbb Z_p$, as they should be, as this symmetry is gauged. Notice that the allowed abelian probe charges $q$ need to satisfy $q=z_e$, where $z_e=e_jp$ is the $N$-ality of the nonabelian line. Thus, we can represent the lines in Eq. (\ref{abelian and nonabelian lines in sunu1modzp}) by the pair $(z_e,q=z_e)$. The theory also possesses a magnetic $U^{(1)}_m(1)$ $1$-form symmetry acting on 't Hooft lines. Let  $z_m=0,1,..,p-1$, and $g$ be the $N$-ality of the nonabelian 't Hooft line and the abelian magnetic charge, respectively. Then, the pairs $(z_e,q=z_e)$ and $(z_m,g)$ must satisfy the Dirac quantization condition $e^{i2\pi (-qg+ z_ez_m/p)}=1$ or $z_ez_m-pqg\in p\mathbb Z$, which gives a constraint on the magnetic charges: $g=\frac{z_m}{p}+\mathbb Z$, i.e., the abelian magnetic charges  can be fractional \cite{Tong:2017oea}. Another way of putting it is that the presence of the Abelian Wilson's lines $W_{j,\,U(1)}^{e_j p}$ demand that the Abelian 't Hooft lines are ${\cal T}_{j,\, U(1)}^{N_j+n_j/p}$, $n_j \in \mathbb Z_p$, $N_j \in \mathbb Z$, such that the electric and magnetic lines are mutually local. The physical states in Hilbert space are taken to be eigenstates of the commuting set of the Hamiltonian, the generators of electric symmetry, and the generators of magnetic symmetry:
\begin{eqnarray}\nonumber
|\psi\rangle_{\scriptsize\mbox{phy}\,,\bm m}&=&|E, p\bm e, \bm n/p+\bm N\rangle_{\bm m}\,, \quad e_j=0,1,..,N/p-1\,,\quad N_j\in\mathbb Z\,\quad  n_j=0,1,..,p-1\,,\\
&&\quad\quad\quad\quad\quad\quad\quad\quad\quad\quad j=1,2,3\,,
\end{eqnarray}
and $\bm m\in \mathbb Z^3$ is the fractional magnetic flux background (or background magnetic twist). Remember that, in principle, $\bm m\in(\mathbb Z\, \mbox{Mod}\, N)^3$; however, we drop the modding by $N$ since this cannot affect the cocycle condition. Notice that we can always activate a $\mathbb Z_N$ magnetic twist since, as emphasized several times,  we use a combination of nonabelian and abelian transition functions. Also, in the special case $p=N$, we should remove the subscript $\bm m$ since, in this case, the  Hilbert space is spanned by eigenstates of the full magnetic $\mathbb Z_N$ fluxes, i.e., $n_j=0,1,...,N-1$.  

The operator $\tilde U_{\mathbb Z_{2T_{\cal R}}, \ell}$ defined in Eq. (\ref{U tilde in SUNU1}) is invariant under both $SU(N)$ and $U(1)$ gauge transformations. However, because we are now gauging $\mathbb Z_p$, the operator must also be invariant under $\mathbb Z_p$ gauge transformations. Let us recall that $T_jt_j$ is the generator of the electric $\mathbb Z_N^{(1)}$ symmetry, and therefore, $\left(T_jt_j\right)^{N/p}$ generates the $\mathbb Z_p$ symmetry, which must be gauged. The action of $\left(T_jt_j\right)^{N/p}$ on $\tilde U_{\mathbb Z_{2T_{\cal R}}, \ell}$  can be read from the first line in Eq. (\ref{sunu1 anomaly}) by applying the operation $N/p$ times:
\begin{eqnarray}
\left(T_jt_j\right)^{N/p} \tilde U_{\mathbb Z_{2T_{\cal R}}, \ell}\left(T_jt_j\right)^{-N/p}= \tilde U_{\mathbb Z_{2T_{\cal R}}, \ell} e^{-2\pi i \ell \left( \f{n_{x}}{p} - \f{n}{p} \f{d_{\rep}}{T_{\rep}} \left(N_x + \f{nn_{x}}{N}\right) \right)}\,.
\end{eqnarray}
This relation shows that for a general $\ell$, $\tilde U_{\mathbb Z_{2T_{\cal R}}, \ell}$ fails to be gauge invariant under a $\mathbb Z_p$ gauge transformation\footnote{In the special case $p=1$, the phase becomes  $e^{2\pi i \ell n \f{d_{\rep}}{T_{\rep}} \left(N_x + \f{nn_{x}}{N} \right)}$, and using Condition (\ref{condition on lx}), the phase trivializes. This shows that this operator is gauge invariant in $SU(N)\times U(1)$ theory, as expected.}.  Being acquainted with the remedy of this problem, we use $\left(T_jt_j\right)^{N/p} \tilde U_{\mathbb Z_{2T_{\cal R}}, \ell}\left(T_jt_j\right)^{-N/p}$ as a building block of a gauge invariant operator by summing over arbitrary copies of the block. The noninvertible operator is then given by
\begin{align}
    \tilde U_{\Z_{2T_{\rep}},\ell} &= \sum_{p_{x}, p_{y},p_{z} \in \Z} (T_{x}t_x)^{\f{Np_{x}}{p}} (T_{y}t_y)^{\f{Np_{y}}{p}} (T_{z}t_z)^{\f{Np_{z}}{p}} U_{\Z_{2T_{\rep}},\ell}  (T_{x}t_x)^{-\f{Np_{x}}{p}} (T_{y}t_y)^{-\f{Np_{y}}{p}} (T_{z}t_z)^{-\f{Np_{z}}{p}} \nonumber \\\nonumber
    & = U_{\Z_{2T_{\rep}},\ell} \sum_{p_{x}, p_{y},p_{z}\in \mathbb Z}  e^{-2\pi i \ell \left( \f{n_{x}}{p} - \f{n}{p} \f{d_{\rep}}{T_{\rep}} \left(N_x + \f{nn_{x}}{N}\right) \right)+(x\rightarrow y)+(x\rightarrow z)}\\
    & = U_{\Z_{2T_{\rep}},\ell} \sum_{l_{x} \in \Z} \d \left( \f{\ell n_{x}}{p} - \f{\ell n}{p} \f{d_{\rep}}{T_{\rep}} \left(N_x + \f{nn_{x}}{N} \right)-l_x\right)\left(\sum_{l_y\in \mathbb Z}...\right)\left(\sum_{l_z\in \mathbb Z}...\right)\,.
 \label{SUU1 one of main results for zp}
\end{align}
The operator $\tilde U_{\Z_{2T_{\rep}},\ell}$ acts invertibly on sectors in Hilbert space that, for $\ell=1$, satisy the condition
\begin{eqnarray}
 l_x=\f{n_{x}}{p} - \f{n}{p} \f{d_{\rep}}{T_{\rep}} \left(N_x + \f{nn_{x}}{N} \right)\in \mathbb Z\,,
 \label{condition for sunu1 zp}
\end{eqnarray}
with identical expressions in the $y$ and $z$ directions. The operator $ \tilde U_{\Z_{2T_{\rep}},\ell}$, as introduced in Eq. (\ref{SUU1 one of main results for zp}), within the context of $SU(N)\times U(1)/\mathbb Z_p$ gauge theory, is a generalization of the operator defined in Eq. (\ref{U tilde in SUNU1}) for the conventional $SU(N)\times U(1)$ theory. Furthermore, Condition (\ref{condition for sunu1 zp}) represents a broader generalization of Condition (\ref{condition on lx}). In the specific scenario where $p=N$ holds, corresponding to the $SU(N)\times U(1)/\mathbb Z_N$ theory, Condition (\ref{condition for sunu1 zp}) precisely mirrors the criterion for the absence of a mixed anomaly between the electric $\mathbb Z_N^{(1)}$ $1$-form global symmetry and the noninvertible chiral symmetry inherent to the $SU(N)\times U(1)$ theory. This correspondence is clear from the first line of Eq. (\ref{sunu1 anomaly}).

The $SU(N) \times U(1)/\mathbb Z_p$ theory exhibits an electric $\mathbb Z_{N/p}^{(1)}$ one-form global symmetry, which is generated by the operators $(T_j t_j)^p$. When introducing a background for this symmetry, we uncover a mixed anomaly between the noninvertible chiral symmetry and the $\mathbb Z_{N/p}^{(1)}$ symmetry. Sandwiching $\tilde U_{\Z_{2T_{\rep}},\ell}$, defined in Eq. (\ref{SUU1 one of main results for zp}), between $(T_x t_x)^p$ and $(T_x t_x)^{-p}$ and using Eqs. (\ref{main Q relation}, \ref{the fractional abelian charge}),  we find
\begin{eqnarray}
\nonumber
(T_x t_x)^p\tilde U_{\Z_{2T_{\rep}},\ell}(T_x t_x)^{-p}&=&\tilde U_{\mathbb Z_{2T_{\cal R}}, \ell} e^{-2\pi i \ell \left( \f{p n_{x}}{N} - \f{pn}{N} \f{d_{\rep}}{T_{\rep}} \left(N_x + \f{nn_{x}}{N}\right) \right)}\\
&=&\tilde U_{\mathbb Z_{2T_{\cal R}}, \ell} e^{-i 2\pi l_x \ell \frac{p^2}{N}}\,,
\label{the anomaly of sunu1zp}
\end{eqnarray}
where we used $l_x$ defined in Eq. (\ref{condition for sunu1 zp}) in going from the first to the second line. When the phase $e^{-i 2\pi l_x \ell \frac{p^2}{N}}$ is nontrivial, it signifies the presence of a degeneracy within the spectrum. Notice that the anomaly phase coincides with the phase in Eq. (\ref{the mixed anomaly zp theories}) if we set $q=N$ in the latter. This should not surprise us since, in this section, we employ the full $\mathbb Z_N$ center symmetry, thanks to gauging $U(1)$. The anomaly in (\ref{the anomaly of sunu1zp}) is valued in $\mathbb Z_{\scriptsize N/\mbox{gcd}(N, p^2l_x)}$ (we take $n_x=n_y=n_z$) indicating a $N/\mbox{gcd}(N, p^2l_x)$-fold degeneracy. The Hilbert space of physical states, which are labeled by $N/p$ distinct electric fluxes, sit in $N/\mbox{gcd}(N, p^2l_x)$ orbits, and a rotation by $\tilde U_{\Z_{2T_{\rep}},\ell=1}$ maps a state with an electric flux $pe_j$ to a state with a flux  $p(e_{\scriptsize j}+\mbox{gcd}(N,p^2l_j)/p )$, i.e., they have the same energy.

\subsection{Examples}

\subsubsection{$SU(4k+2)\times U(1)/\mathbb Z_p$ with $2$-index antisymmetric fermions}

$SU(4k+2)\times U(1)$ theory with a single $2$-index anti-symmetric Dirac fermion was considered in \cite{Anber:2023pny}. Here, we study this theory when we gauge a $\mathbb Z_p\subseteq \mathbb Z_N$ subgroup of the center. Numerical scans reveal that condition (\ref{condition for sunu1 zp}) is always satisfied for specific values of $n_{x}$ and $N_x$.  Also, the anomaly (\ref{the anomaly of sunu1zp}) is trivial unless both  $p$ and $l_x$ are odd; then, the anomaly is valued in $\mathbb Z_2$. The Hilbert space is spanned by the physical states
\begin{eqnarray}\nonumber
|\psi\rangle_{\scriptsize\mbox{phy}\,,\bm m}&=&|E, p\bm e, \bm n/p+\bm N\rangle_{\bm m}\,,\\\nonumber
&& \quad e_j=0,1,..,(4k+2)/p-1\,,\quad N_j\in\mathbb Z\,,\quad  n_j=0,1,..,p-1\,,
\quad  j=1,2,3\,,\\
\end{eqnarray}
and the anomaly means that the states live in two orbits such that $|E, p\bm e, \bm n/p+\bm N\rangle_{\bm m}$, $|E, p(\bm e+\mbox{gcd}(N,p^2 \bm l)/p), \bm n/p+\bm N\rangle_{\bm m}$,  $|E, p(\bm e+2\mbox{gcd}(N,p^2 \bm l)/p), \bm n/p+\bm N\rangle_{\bm m}$, etc. have the same energy (we take $n_x=n_y=n_z$).

\subsubsection{The Standard Model}

The methods presented in this paper provide a systematic means to find noninvertible symmetries in any given gauge theory. As an important application, we employ our approach to search for noninvertible symmetries in the nongravitational sector of the Standard Model (SM). SM is based on $su(3)\times su(2)\times u(1)$ Lie algebra. Yet, the faithful gauge group, i.e., the global structure of the group, is to be uncovered.  The matter content and charges under the gauge and global symmetries are displayed in Table \ref{charges of SM}, and all fermions are taken to be left-handed Weyls.
\begin{table}
\begin{equation}\nonumber
\begin{array}{|c|ccc|cc|}\hline\text{field}&SU(3)& SU(2)&U(1)&U(1)_B& U(1)_L \cr \hline
q_L&\square&\square&1&{1\over 3}& 0\cr
l_L &\bf{1}&\square&-3&0&1 \cr
\tilde{e}_R&\bf{1}&\bf{1}&6&0&-1\cr
\tilde{u}_R&\overline\square&\bf{1}&-4&-{1\over 3}&0\cr
\tilde{d}_R&\overline\square&\bf{1}&2&-{1\over 3}&0\cr
h&\bf{1}&\square&3&0&0\cr\hline
\end{array}
\end{equation}
\caption{Matter content and charges of  SM: $q_L$ and $l_L$ are the quark and lepton doublets, $\tilde{e}_R, \tilde{u}_R, \tilde{d}_R$ are the electron and up and down quarks singlets, while $h$ is the Higgs doublet. Notice that we take the hyper $U(1)$ charges to be integers, while the matter content has the standard charges under the baryon number $U(1)_B$ and lepton number $U(1)_L$ symmetries.}
\label{charges of SM}
\end{table}
The anomalies associated with the $U(1)_{B}$ and $U(1)_{L}$ symmetries are given by:
$U(1)_{B} \times [SU(2)]^{2} = U(1)_{L} \times [SU(2)]^{2} = 1$, 
$U(1)_{B} \times [SU(3)]^{2} = U(1)_{L} \times [SU(3)]^{2} = 0$,
$U(1)_{B} \times [U(1)]^{2}=U(1)_{L} \times [U(1)]^{2}=-18$\,.
Thus, we see that $U(1)_{B-L}$ symmetry is anomaly-free symmetry (we neglect gravity in this context). Under a $U(1)_{B+L}$ rotation, the path integral picks up an ABJ phase
\begin{equation}
    \exp \left( i \a \cdot N_f(2 c_{2}(F) - 36 c_{2}(f) ) \right)\,, 
\end{equation}
where $N_f$ is the number of families, $c_{2}(F)$ is the second Chern class for $SU(2)$ and $c_{2}(f)$ is the second Chern class for $U(1)$. The ABJ anomaly breaks the $U(1)_{B+L}$ down to a $\Z^{B+L}_{\scriptsize\mbox{gcd}(2,36) N_f}=\Z^{B+L}_{2N_f}$ symmetry. Notice that $SU(3)$ does not play a role in the ABJ anomaly.

The matter content is consistent with the existence of an electric  $\mathbb Z_6^{(1)}$ $1$-form global symmetry \cite{Tong:2017oea,Anber:2021upc}. The cocycle conditions satisfied by SM on $\mathbb T^4$ with a gauged $\Z_{6}^{(1)}$ are given by \cite{Anber:2021upc}: 
\begin{align}
    \Omega_{(3)\mu}(x^{\nu} = L^{\nu}) \Omega_{(3)\nu} (x^{\mu} = 0) & = e^{2\pi i \f{n^{(3)}_{\mu\nu}}{3}} \Omega_{(3)\nu}(x^{\mu} = L^{\mu}) \Omega_{(3)\mu} (x^{\nu} = 0)\,, \nonumber \\
    \Omega_{(2)\mu}(x^{\nu} = L^{\nu}) \Omega_{(2)\nu} (x^{\mu} = 0) & = e^{2\pi i \f{n^{(2)}_{\mu\nu}}{2}} \Omega_{(2)\nu}(x^{\mu} = L^{\mu}) \Omega_{(2)\mu} (x^{\nu} = 0)\,, \\
    \omega_{(1)\mu}(x^{\nu} = L^{\nu}) \omega_{(1)\nu} (x^{\mu} = 0) & = e^{-2\pi i (\f{n^{(3)}_{\mu\nu}}{3} + \f{n^{(2)}_{\mu\nu}}{2} )} \omega_{(1)\nu}(x^{\mu} = L^{\mu}) \omega_{(1)\mu} (x^{\nu} = 0)\,. \nonumber
\end{align}
$\Omega_{(i)}$, $i=2,3$, and $\omega_{(1)}$ are the transition functions of the gauge bundles, $n^{(i)}_{\mu\nu}$ are the 't Hooft twists, and the superscript/subscript $(i) = (3), (2), (1)$ denote the condition for the $SU(3), SU(2), U(1)$ gauge groups respectively. The electric $\Z_{6}^{(1)}$ symmetry is generated by a combinations of the $SU(3)$ center, $T^{(3)}_{j}$, the $SU(2)$ center, $T^{(2)}_{j}$, and the $U(1)$ center $t_j$, such that the full $\Z_{6}^{(1)}$  symmetry generator is given by $T^{(3)}_{j} T^{(2)}_{j} t_{j}$, $j=x,y,z$.

The anomalous $U(1)_{B+L}$ current conservation law is given by
\begin{equation}
    \del_{\mu} j^{\mu}_{B+L} - 2N_f \del_{\mu} K^{\mu}_{SU(2)} (A) + \f{36N_f}{8\pi^{2}} \epsilon_{\mu \nu \l \s} \del^{\mu} a^{\nu} \del^{\l} a^{\s}= 0\,,
\end{equation}
where $K^{\mu}_{SU(2)}$ is the $SU(2)$ topological current. The corresponding unbroekn  $\Z^{B+L}_{2N_f}$ symmetry operator on $\mathbb T^3$  is given by:
\begin{eqnarray}
U_{\mathbb Z_{2N_f}, \ell} =\exp\left[i \frac{2\pi \ell}{2N_f} Q_5 \right]\,,
\end{eqnarray}
where the conserved charge $Q_5$ is given by (here we turn on a $\mathbb Z_6$ magnetic twist)
\begin{eqnarray}\label{q5 SM}\nonumber
Q_5 &=& \int_{\mathbb T^3} d^3 x \left[  j_{B+L}^0 - 2 N_f K^{0}_{SU(2)}(A) + \frac{36 N_f}{8\pi^2}\epsilon^{ijk}  a_i \partial_j a_k \right]\\
&&-  { 18 N_f \over 4 \pi} (N_{z} + {1\over 6} n_{z})\left[ \int\limits_{0}^{L_y} {d y \over L_y} \int\limits_{0}^{L_z} d z a_z(x=0,y,z) +\int\limits_{0}^{L_x} {d x \over L_x} \int\limits_{0}^{L_z} d z a_z(x,y=0,z)  \right]  \nonumber\\
&&+ \sum\limits_{\scriptsize\mbox{cyclic}}  (x \rightarrow y \rightarrow z \rightarrow x)\,.
\end{eqnarray}
Under a $U(1)$ gauge transformation, $U_{\mathbb Z_{2N_f}, \ell}$ transforms as
\begin{eqnarray}
U_{\mathbb Z_{2N_f}, \ell} \longrightarrow U_{\mathbb Z_{2N_f}, \ell} e^{-i2\pi\left(\f{ 18 \ell N_f}{N_f} \left( N_{x} + \f{n_{x}}{6}\right)\right)+(x\rightarrow y)+(x\rightarrow z)}=U_{\mathbb Z_{2N_f}, \ell} \,.
\end{eqnarray}
Therefore, $U_{\mathbb Z_{2N_f}, \ell}$ is $U(1)$ gauge invariant, as required. Further, we examine  $U_{\mathbb Z_{2N_f}, \ell}$ after gauging the electric $\mathbb Z_6^{(1)}$ $1$-form center by sandwiching $U_{\mathbb Z_{2N_f}, \ell}$ between its generators (this is a generalization of Eq. (\ref{sunu1 anomaly})):
\begin{eqnarray}
\nonumber
T_x^{(3)}T_x^{(2)}t_xU_{\mathbb Z_{2N_f}, \ell}\left(T_x^{(3)}T_x^{(2)}t_x\right)^{-1}&=&\underbrace{e^{-i \frac{2\pi \ell(2N_f)}{2N_f} \frac{n_x^{(2)}}{2}}}_{\scriptsize \mbox{from}\,K^{0}_{SU(2)}(A) }\underbrace{e^{i 2\pi \ell \frac{36 N_f}{2N_f}\left(\frac{1}{6}\right)\left(N_x+\frac{n_x^{(2)}}{2}+\frac{n_x^{(3)}}{3}\right)}}_{\scriptsize \mbox{from}\,\epsilon^{ijk}  a_i \partial_j a_k}U_{\mathbb Z_{2N_f}, \ell}\\
&=&U_{\mathbb Z_{2N_f}, \ell}\,.
\end{eqnarray}
We used Eq. (\ref{the manipulations of U}), setting $k_x=m_x=1$, to find the first exponent. The second exponent is found by applying Eq. (\ref{the fractional abelian charge}) and using $n=1$, $N=6$. Here, $n_x^{(2)}$, $n_x^{(3)}$, and $N_x$ are the $SU(2)$ and $SU(3)$ fractional twists and $U(1)$ integral magnetic flux, respectively.
 This analysis shows that SM does not possess noninvertible symmetries in its nongravitational sectors. Our findings are consistent with \cite{Putrov:2023jqi}.

\section{Coupling gauge theories to axions and noninvertible symmetries}
\label{Coupling gauge theories to axions and noninvertible symmetries}

In this section, we introduce axions into the game, taking $\mathbb T^4$ to be larger than any scale in the theory. To be specific, we take $SU(N)/\mathbb Z_p$ or $SU(N)\times U(1)/\mathbb Z_p$  gauge theories of the previous sections and follow the setup of \cite{Anber:2020xfk} by adding a complex scalar $\Phi$ that is neutral under the gauge groups but couples to the fermions. Thus, we add the following terms to the Lagrangian:
\begin{eqnarray}
{\cal L}\supset |\partial_\mu \Phi|^2+V(\Phi)+y \Phi \tilde \psi \psi+\mbox{h.c.}\,,
\end{eqnarray}
where $\psi, \tilde \psi$ are two left-handed Weyl fermions in representations ${\cal R}$ and its complex conjugate $\bar{\cal R}$, respectively, and $y$ is a Yukawa coupling.  The potential of the complex field is $V(\Phi)=\lambda(|\Phi|^2-v^2)^2$, where $\lambda$ is ${\cal O}(1)$ dimensionless parameter. We take the scalar field v.e.v. $v\gg \Lambda$, where $\Lambda$ the strong scale of the gauge sector.  We shall pretend that we did not know about the noninvertible symmetries or how to construct them, and let us see if we can reproduce them in the IR.

 Let us first consider the $SU(N)$ gauge theory before gauging $U(1)$ and the electric $\mathbb Z_p^{(1)}$ symmetry.  Under $\mathbb Z_{2T_{\cal R}}^\chi$ and $U(1)$ baryon number, the different fields transform as 
\begin{eqnarray}
\nonumber
\mathbb Z_{2T_{\cal R}}^\chi&:&\quad \Phi\longrightarrow e^{i\frac{-2\pi}{T_{\cal R}}} \Phi\,, \psi\longrightarrow e^{i\frac{2\pi}{2T_{\cal R}}} \psi\,, \tilde \psi\longrightarrow e^{i\frac{2\pi}{2T_{\cal R}}} \tilde\psi\,,\\
U(1)&:&\quad \Phi\longrightarrow \Phi\,, \psi\longrightarrow e^{i\alpha}\psi\,, \tilde\psi\longrightarrow e^{-i\alpha}\tilde\psi\,, \alpha\in [0,2\pi)\,.
\end{eqnarray}
If we write $\Phi$ as $\Phi=\rho e^{i\varphi}$, where $\varphi$ is the axion, then $\varphi$ transforms under $\mathbb Z^\chi_{2T_{\cal R}}$ as
\begin{eqnarray}
\varphi\longrightarrow \varphi-\frac{2\pi}{T_{\cal R}}\,
\label{DSS}
\end{eqnarray}
and notice that the axion is inert under the $\mathbb Z_2^F$ fermion number subgroup of  $\mathbb Z_{2T_{\cal R}}^\chi$ .

Next, we consider $SU(N)/\mathbb Z_p$ or $SU(N)\times U(1)/\mathbb Z_p$ gauge theories with axions. Flowing to an energy scale below $v$, the radial degree of freedom $\rho$ freezes in, i.e., we set $\rho=v$, and the fermions acquire a mass $\sim yv$ and decouple.  What remains is the light degree of freedom, the axion $\varphi$.  However, the axion should reproduce all the UV anomalies.  Thus, we can write the following IR effective Lagrangian of $\varphi$:
 \begin{eqnarray}
 {\cal L}_{\varphi}=v^2\left(\partial_\mu \varphi\right)^2+ T_{\cal R} \varphi\frac{\mbox{tr}(F\wedge F)}{8\pi^2}+d_{\cal R} \varphi\frac{f\wedge f}{8\pi^2}\,.
 \end{eqnarray}
Variation of ${\cal L}_\varphi$  w.r.t $\varphi$ produces the anomalous  current conservation law:
\begin{eqnarray}
\partial_\mu  j^\mu_{(\varphi)}-T_{\cal R}\partial_\mu K^\mu(A)-\frac{d_{\cal R}}{8\pi^2}\epsilon_{\mu\nu\lambda\sigma}\partial^\mu a^\nu\partial^\lambda  a^\sigma=0\,,
\end{eqnarray}
where $ j^\mu_{(\varphi)}=v^2\partial^\mu \varphi$. This is exactly the anomalous current conservation law we had previously, now written down for the axion current. Therefore, everything we said in the previous sections applies here. In particular, we can define an operator of the $\mathbb Z_{2T_{\cal R}}^\chi$ symmetry as:
\begin{eqnarray}
{\cal U}_{\mathbb Z_{2T_{\cal R}}\,,\ell}= \exp\left[i \frac{2\pi \ell}{T_{\cal R}} \int_{\mathbb T^3} ( j_{(\varphi)}^0-T_{\cal R} K^0( A))+...  \right]\,,
\end{eqnarray}
where the dots denote the contribution from the $U(1)$ gauge field (see Eq. (\ref{q5})). We used a calligraphic letter for the operator to emphasize that it is constructed in the IR. Yet, all the anomalies and failure of invariance under gauge symmetries that lead to the noninvertibility of the UV operators apply here as well. Thus, similar to what we did before, we can construct the noninvertible operator $\tilde {\cal U}_{\mathbb Z_{2T_{\cal R}}\,,\ell}$, which implements the noninvertible symmetry $\tilde{\mathbb Z}_{2T_{\cal R}}^\chi$ in the IR. Such operators shall project onto magnetic sectors and also exhibit mixed anomalies with the global $1$-form electric center symmetry, exactly as we discussed previously.

 It was pointed out in \cite{Cordova:2023her} that $SU(N)/\mathbb Z_p$ theories with axions have noninvertible symmetries. However, our construction shows that such a conclusion is not general and depends on the UV completion.
Consider two theories $SU(4k)/\mathbb Z_2$ and $SU(4k+2)/\mathbb Z_2$ with a Dirac fermion in the $2$-index antisymmetric representation and coupled to a complex scalar field $\Phi$ as above. As we flow to the IR, we can construct the operators corresponding to the chiral symmetries.  We discussed in Section \ref{YM theories with fermions in the 2index antisymm} that $SU(4k)/\mathbb Z_2$ theory does not exhibit an anomaly between its chiral symmetry and the $1$-form symmetry of the corresponding $SU(4k)$ theory, and hence, the chiral symmetry operator is invertible. Therefore, an axion domain wall (DW), implemented by the action of $\tilde {\cal U}_{\mathbb Z_{8k-4}\,,\ell}$, will not be dressed with TQFT degrees of freedom. On the contrary, $SU(4k+2)/\mathbb Z_2$ exhibits an anomaly between its chiral symmetry and the $1$-form center of the corresponding $SU(4k+2)$ theory, and thus, the minimal chiral symmetry operator $\tilde {\cal U}_{\mathbb Z_{8k}\,,\ell=1}$ is noninvertible. The axion DW implemented by  $\tilde {\cal U}_{\mathbb Z_{8k}\,,\ell=1}$ must be dressed with a fractional quantum Hall TQFT. 

We may also consider axions in $SU(N)\times U(1)/\mathbb Z_p$ theory of Section \ref{SUNTIMES U1MODZP}. Everything we said there is transcendent to the IR axion domain walls. In particular, for $p=1$, the operator $\tilde {\cal U}_{\mathbb Z_{2T_{\cal R}}\,,\ell=1}$ destructs  the Hilbert space sectors characterized by vanishing fractional $\bm n=0$ and integral magnetic fluxes $\bm N \notin  T_{\cal R}  \mathbb Z^3/\mbox{gcd}(T_{\cal R}, d_{\cal R})$. It will be interesting to examine what happens to the axion domain walls of this theory as we place them in such an external magnetic field.

{\bf {\flushleft{Acknowledgments:}}} We would like to thank Erich Poppitz and Tin Sulejmanpasic for various illuminating discussions. We also thank Erich Poppitz for comments on the manuscript.   This work is supported by STFC through grant ST/T000708/1. 

  \bibliography{Refnoninv-gauging.bib}
 
  \bibliographystyle{JHEP}

  \end{document}